\documentclass[11pt]{article}
\usepackage{float}
\usepackage[page,header]{appendix}
\usepackage{titletoc}

\usepackage{natbib}

\RequirePackage[OT1]{fontenc}
\RequirePackage{amsthm,amsmath}
\RequirePackage[colorlinks,citecolor=blue,urlcolor=blue]{hyperref}
\RequirePackage{hypernat}
\usepackage{amssymb,tabularx,multicol,multirow,booktabs}
\usepackage{mhequ}
\usepackage{fancyvrb}

\usepackage{bm}
\usepackage{bbold}
\usepackage[table]{xcolor}

\usepackage{caption}
\usepackage{subcaption}
\providecommand{\nopunct}{\spacefactor \@nopunct}
\def\@nopunctsfcode{1007}

\usepackage{tikz}
\usetikzlibrary{calc, shapes, backgrounds}
\usepackage{subcaption}
\usepackage[labelformat=parens,labelsep=quad,skip=3pt, font=small, labelfont=bf]{caption}
\usepackage{graphicx}
\usepackage[export]{adjustbox}

\newtheorem{theorem}{Theorem}[section]

\newtheorem{prop}[theorem]{Proposition}

\newtheorem{assumption}[theorem]{Assumption}

\usepackage{color}
\usepackage{amssymb}
\usepackage{comment}
\usepackage{pdfpages}
\usepackage{dcolumn}

\numberwithin{equation}{section}

\usepackage{scalerel,stackengine}
\stackMath

\newcommand\numberthis{\addtocounter{equation}{1}\tag{\theequation}}

\def\b1{\boldsymbol{1}}

\newcommand{\pr}[1]{\mathbb{P} \! \left \{ #1 \right \}}
\newcommand{\EE}[1]{\E \left [ #1 \right ]}

\newcommand{\E}{\mathbb{E}}

\newcommand{\yhat}{\widehat{Y}}
\newcommand{\ybar}{\overline{Y}}

\DeclareMathOperator{\Cov}{Cov}
\DeclareMathOperator{\Var}{Var}

\newcommand\reallywidehat[1]{%
\savestack{\tmpbox}{\stretchto{%
  \scaleto{%
    \scalerel*[\widthof{\ensuremath{#1}}]{\kern-.6pt\bigwedge\kern-.6pt}%
    {\rule[-\textheight/2]{1ex}{\textheight}}
  }{\textheight}%
}{0.5ex}}%
\stackon[1pt]{#1}{\tmpbox}%
}

\definecolor{mygrey}{gray}{0.4}

\newcommand{\cen}{\centering}

\newcommand{\ITT}{\text{ITT}}
\newcommand{\LATE}{\text{LATE}}
\newcommand{\obs}{\text{obs}}

\newcolumntype{M}[1]{>{\centering\arraybackslash}m{#1}}
\newcolumntype{N}{@{}m{0pt}@{}}

\def \be{\begin{equs}}
\def \ee{\end{equs}}

\def \E{\mathbb{E}}

\begin{document}

\startcontents[sections]

\title{Identifying and Estimating Principal Causal Effects in Multi-site Trials\thanks{We gratefully acknowledge funding from the Spencer Foundation through a grant entitled ``Using Emerging Methods with Existing Data from Multi-site Trials to Learn About and From Variation in Educational Program Effects'' and from the Institute for Education Sciences, U.S. Department of Education, through Grant \#R305D150040. The opinions expressed are those of the authors and do not represent views of the Institute or the U.S. Department of Education.}}

\author{Lo-Hua Yuan\\Harvard University \and Avi Feller\\UC Berkeley \and Luke W. Miratrix\\Harvard GSE}

\maketitle

\begin{abstract}
Randomized trials are often conducted with separate randomizations across multiple sites such as schools, voting districts, or hospitals. These sites can differ in important ways, including the site's implementation, local conditions, and the composition of individuals. An important question in practice is whether---and under what assumptions---researchers can leverage this cross-site variation to learn more about the intervention. We address these questions in the principal stratification framework, which describes causal effects for subgroups defined by post-treatment quantities. We show that researchers can estimate certain principal causal effects via the multi-site design if they are willing to impose the strong assumption that the site-specific effects are uncorrelated with the site-specific distribution of stratum membership.
We motivate this approach with a multi-site trial of the Early College High School Initiative, a unique secondary education program with the goal of increasing high school graduation rates and college enrollment. Our analyses corroborate previous studies suggesting that the initiative had positive effects for students who would have otherwise attended a low-quality high school, although power is limited.
\end{abstract}


\section{Introduction}

Randomized trials are often conducted at multiple physical sites, with separate randomizations across, for example, schools, voting districts, or hospitals~\citep{raudenbush2015learning}. These sites can differ in important ways, including the site's implementation quality, local conditions, and the composition of individuals. Intuitively, researchers should be able to leverage such differences across sites to learn more about the intervention. For instance, if impacts are systematically larger at sites with higher student attendance, what can we conclude about dosage effects? More broadly, what questions can researchers answer using this approach and what assumptions are required?

This paper explores the use of cross-site variation to estimate causal effects defined by individual-level post-treatment behavior. Our motivating example is a randomized evaluation of an alternative high school program in North Carolina, known as Early College High Schools~\citep[ECHS;][]{edmunds2012expanding}. ECHS is an innovative approach that aims to increase college readiness and college completion rates among students typically under-represented in post-secondary education.~\citet{edmunds2017smoothing} find meaningful, positive impacts on a range of key academic outcomes, including ninth-grade success, high school graduation, and college enrollment. These positive results raise additional questions about expanding the program. In particular, is it more effective for certain types of students or in certain settings?

Our analysis focuses on the quality of the school each student would attend in the absence of the program. 
In general, we expect to see larger impacts of ECHS for students who would otherwise attend low-quality public schools than for those who would otherwise attend high-quality public schools. 
The goal is to assess whether this indeed holds in practice, which would help guide the expansion of the program. 
We make this question precise via the {\em principal stratification} framework of~\citet{frangakis02} and define subgroups, known as principal strata, determined by each student's school quality in both the observed treatment condition and the counterfactual condition. 
While membership in these endogenous subgroups is only partially observed, the corresponding causal effects are nonetheless well defined.

Although principal stratification is a powerful framework for defining causal effects of interest, estimating these impacts can be elusive~\citep{Page2015}. 
In the context of multi-site trials, we show that estimation is possible via a {\em zero correlation assumption}: the site-specific distribution of principal strata (e.g., the proportion of Compliers) is uncorrelated with the site-specific impacts for these principal strata. 
This is a very strong assumption, roughly implying that the interaction between randomization and site indicator functions as a  ``second instrument'' (the first being treatment randomization) that is predictive of principal stratum membership, but is uncorrelated with the treatment impact within any stratum. As we argue, multi-site trials differ from more general stratified randomized trials because we can appeal to a (super) population of sites. Thus, rather than assume that certain quantities are constant and equal to zero for all sites, we can instead assume that these quantities equal zero \emph{on average} across sites~\citep[see][]{kolesar2015identification}. 
We describe this zero-correlation assumption in the context of principal stratification in the ECHS study.
We also address estimation and discuss the weaker assumption that zero correlation only holds conditional on a set of auxiliary covariates.

To the best of our knowledge, this is the first paper that brings together the otherwise disparate literatures of multi-site trials and covariate restrictions for principal stratification. 
We mention several highly relevant papers, and explore the connections in more depth in Section~\ref{sec:ties}. 
First,~\citet{reardon2013_msmm_iv} outline nine assumptions required to estimate mediation (rather than principal stratification) effects via cross-site variation~\citep[see also][]{reardon2014bias, raudenbush2015learning}. 
Second,~\citet{kolesar2015identification} explore related questions from an econometric perspective and consider estimation with ``many invalid instruments.'' Both~\citet{reardon2013_msmm_iv} and~\citet{kolesar2015identification} impose a zero correlation assumption very similar to the one we explore here, though our setup gives researchers greater flexibility by requiring fewer necessary conditions for identification and estimation. 
Third,~\citet{Jiang2016} discuss identifying principal causal effects by leveraging results from multiple studies. They impose the much stronger assumption that these effects are constant (``homogeneous'') across studies~\citep[see also][for additional discussion]{KlineWalters2014}. 
Many other papers impose restrictions on covariates to identify principal causal effects, including~\citet{Jo:2002cw},~\citet{Peck2003},~\citet{ding2011identifiability} and~\citet{mealli2016concentration}. 
Finally,~\citet{miratrix2017bounding} investigate the same substantive question that we explore here, but use covariates to sharpen bounds rather than to obtain point estimates.

The paper proceeds as follows. 
Section~\ref{sec:echs} describes the multi-site Early College High School study.
Section~\ref{sec:setup} formulates the principal strata and associated estimands for ECHS.
Section~\ref{sec:identification_general} gives the key methodological results, including identification and estimation.
Section~\ref{sec:aux_covariates} extends these results to incorporate auxiliary covariates.
Section~\ref{sec:echs_results} presents the results for the ECHS study.
Sections~\ref{sec:ties} and~\ref{sec:conclusion} discuss connections to other methods and conclude. 
The supplementary materials contain implementation details, an extensive simulation study, and additional discussion of other methods, especially ASPES~\citep{Peck2003}. 

\section{Early College High Schools} \label{sec:echs}


The Early College High School (ECHS) Initiative was launched in 2002 with support from the Bill and Melinda Gates Foundation. The program partners small, autonomous public high schools with two- or four-year colleges to give students the opportunity to earn an associate's degree or up to two years of transferable college credit, as well as a high school diploma.
Early Colleges are designed to increase college readiness and graduation rates by exposing high school students to college-style courses, building students' confidence in their ability to succeed in a college environment, and lessening the financial burden of college by giving students the option to earn college credits while still in high school.
These programs are targeted at individuals generally under-represented in college, including low income, first generation, and minority students.
Early College programs were oversubscribed at some sites, which then allocated slots to applicants randomly, creating a de-facto randomized trial.

We analyze data from the Evaluation of Early College High Schools in North Carolina~\citep{Edmunds2010}. 
This study tracked a sample of 4,004 students who began ninth grade between 2005 and 2010 and who entered in one of 44 lotteries to gain entry into one of 19 different Early College programs. These ECHS programs are spread across the state, such that it was only feasible for a student to enter into a single lottery.
Within each lottery, students were randomized either to receive or not receive an offer to attend an ECHS. 
Following~\citet{miratrix2017bounding}, we limit our analytic set to students who could be linked to the North Carolina Department of Instruction (NCDPI) databank, had school enrollment data in ninth grade, and had transcript data or End of Course exam data from NCDPI. 
We subset our sample to students whose ninth grade school was within 20 miles of their eighth grade school, under the assumption that a large distance between a student's middle and high schools indicates that the student moved between eighth and ninth grade, and was therefore effectively dropped from the trial.
We also exclude students for whom we do not have complete information on race, gender, free or reduced-price lunch eligibility, first generation college student status, and eighth grade math and reading scores.
Finally, to avoid unnecessary technical complications in the main text, we exclude the six lotteries that have no variability in our outcome measure of interest. We report the same analysis with all 44 lotteries in the supplement, which yields nearly identical conclusions.

Given these inclusion criteria, our final ECHS analysis sample consists of 3,477 students ($N_t = 2,021$, $N_c = 1,456$) across 38 lotteries in 18 ECHS schools, each with up to 6 cohorts. 
Throughout, we use the term `site' to denote a specific lottery rather than a specific school. 
A key reason for this choice is that the proportion of principal strata can vary meaningfully within a school year to year, which complicates school-level analyses. 



\paragraph{Outcomes.} \label{sec:outcomes}
The North Carolina ECHS data set contains a battery of outcome measures. 
Our outcome of interest is a binary indicator of whether a student is ``on track'' to complete the Future-Ready Core Graduation Requirements set by the state of North Carolina at the end of ninth grade. 
This measure is based on compelling descriptive evidence that students who do well in ninth grade are more likely to excel in and graduate from high school~\citep{allensworth2005graduation}.\footnote{Details of the Future-Ready Core's requirements for math and English language reading and writing are at \url{http://www.dpi.state.nc.us/docs/gradrequirements/resources/gradchecklists.pdf}.}

\paragraph{Covariates.}  \label{sec:covariates}
Student baseline covariates include race, gender, free or reduced-price lunch eligibility, first generation college student status, and standardized eighth grade math and reading scores. 
Table 1
in the supplementary materials shows balance checks, stratified by lottery. 
Early College High Schools target students who would traditionally not enroll in college, and several schools in the study gave priority to groups underrepresented in higher education. 
As such, the ECHS sample is relatively disadvantaged, with around half of all students in the lottery eligible for free or reduced-price lunch. 
We also see slight imbalances in racial categories, with the treatment group comprised of more Black/African American students than the control group.
We do not detect imbalance in any of the other baseline covariates. 


\paragraph{Student sampling weights.} \label{sec:weights}
In the ECHS study, students had unequal but known probabilities of winning a lottery.
Some lotteries were more selective overall. Some lotteries gave certain students higher chances of a slot for equity reasons.
All the calculations we perform on the ECHS data set use student-level sampling weights that reflect each student's probability of entering and winning a lottery based on demographics and other factors.
In particular, we apply the same H{\`a}jek estimator sample weighting approach discussed and used by \citet{miratrix2017bounding}.

\paragraph{School quality.}
We label each school in the North Carolina Early College Study as one of three school types: high-quality public high school, low-quality public high school, or Early College High School. 
The high- and low- quality ratings are based on a composite of school-level measures, including achievement metrics, growth, and adequate yearly progress, as tracked by a centralized State of North Carolina school-report-card system. 
Schools classified by the state as ``priority schools'', ``low performing schools'', and ``schools receiving no recognition'' are categorized as low-quality schools. ``Schools making high growth'', ``schools making expected growth'', ``honor schools of excellence'', ``schools of excellence'', and ``schools of progress'' are classified as high-quality schools.\footnote{See~\url{http://www.ncpublicschools.org/docs/accountability/reporting/abc/2005-06/execsumm.html} for classification details.}
While the state also rates Early Colleges as either low- or high-quality, we treat ECHSs as their own quality category because an ECHS operates on principles that are distinct from a traditional public high school and provides students with a unique education environment that may not be captured by standard school rating measures.

Table~\ref{tab:schltype_dist} shows the distribution of ninth grade students in our data set across these three school types. 
In the treatment group, 85.4\% of students attended an ECHS; 2.4\% attended a high-quality school; 12.3\% attended a low-quality school. 
In the control group, only 2.7\% percent were able to cross over and register in an ECHS; 12.4\% attended a high-quality school; 85\% attended a low-quality school.

\begin{table}[bt]
\center
\caption{Distribution of high school type by treatment status}
\label{tab:schltype_dist}
\begin{tabular}{lcc}
\hline
 \multirow{2}{*}{School type}  & Treatment & Control \\
 & ($N_t = 2,021$) & ($N_c = 1,456$) \\
\hline
Early College HS ($e$) &	85.4\% &	2.7\% \\
High-Quality Public HS ($hq$) &	2.4\%	& 12.4\% \\
Low-Quality Public HS ($lq$) &	12.3\%	& 85.0\% \\
\hline
\end{tabular}
\end{table}

\section{Setup and estimands} \label{sec:setup}

We now describe the setup and estimands for the ECHS study using the principal stratification framework. 
%
Let $Z_i$ be the treatment indicator for whether student $i$ is randomly assigned to the active intervention, i.e., wins the lottery and is invited to enroll in an ECHS. 
Let $Y_i^{obs}$ denote student $i$'s observed outcome, i.e., the student's on-track status at the end of her ninth grade academic year. 
We assume randomization was valid within each lottery and that lotteries are independent.
We also invoke SUTVA~\citep{rubin1980}, assuming that there is no interference between units and that there is one version of each treatment level; this precludes murky communication of whether someone wins the lottery and is invited to enroll in an ECHS. 
With these assumptions, we can then write down the potential outcomes for student $i$ as $Y_i(1)$ and $Y_i(0)$, which are student $i$'s on-track status depending on whether or not she receives an Early College enrollment offer.
Her observed on-track status is $Y_i^{\obs} = Z_iY_i(1) + (1-Z_i)Y_i(0)$.

Given this setup, the overall Intent-to-Treat (ITT) effect is therefore
$$ \text{Overall ITT } = \E[Y_i(1) - Y_i(0)],$$
the average impact of the ECHS enrollment offer on students' on-track status. 
For ease of exposition, we initially regard expectations and probabilities as being taken over a super-population of individuals, with individuals from a specific lottery as a random sample of this super-population.
We discuss a corresponding super-population of sites in Section~\ref{sec:identification_general}.

We can now go beyond the overall impact of randomization using the principal stratification framework.
Let $D_i(z) \in \{e, \; lq, \; hq\}$ denote the quality of school a student would attend if assigned to treatment level $Z_i=z$, where $e$, $lq$, and $hq$ are abbreviations for ECHS, low-quality, and high-quality, respectively. 
We now define our principal strata $S_i$ by the pair of school types a student would attend if assigned to treatment, $D_i(1)$, and if assigned to control, $D_i(0)$. 

\begin{table}[bt]
  \caption[Diagram of assumed principal strata in ECHS study]{The nine possible principal strata in the ECHS study. We assume that strata {\em (A)} - {\em(D)} do not exist, leaving five principal strata. The two highlighted cells indicate the strata of interest.}
   \label{tab:echs_table_5strata}
	\begin{tabular}{c|M{20mm}|M{27mm}|M{27mm}|M{27mm}|N}
		\multicolumn{2}{c}{} & \multicolumn{3}{c}{\cen \ \ \ \bf No ECHS offer ($Z_i=0$)} &
		\\[10pt] \cline{3-5}
		\multicolumn{1}{c}{} & & \multicolumn{1}{c|}{$D_i(0)=e$} & \multicolumn{1}{c|}{$D_i(0)=lq$} & \multicolumn{1}{c|}{$D_i(0)=hq$} &
		\\[20pt] \cline{2-5}
		\multirow{4}{*}{\rotatebox[origin=c]{0}{\parbox[c]{1.5cm}{\cen \bf ECHS offer ($Z_i=1$)}} } & $D_i(1)=e$ & \textbf{ECHS Always Taker} & \cellcolor{lightgray} \textbf{Low-Quality Complier} & \cellcolor{lightgray} \textbf{High-Quality Complier} &
		\\[20pt] \cline{2-5}
		& $D_i(1)=lq$ & {\em (A)} & \textbf{Low-Quality Always Taker} & {\em (C)} &
		\\[20pt] \cline{2-5}
		& $D_i(1)=hq$ &{\em (B)} & {\em (D)} & \textbf{High-Quality Always Taker} &
		\\[20pt] \cline{2-5}
	\end{tabular}
\end{table}

Table~\ref{tab:echs_table_5strata} shows the $3^2=9$ possible principal strata; rows indicate school type for students when assigned to treatment and columns indicate school type when assigned to control. 
The analysis becomes unwieldy without restrictions on the possible principal strata~\citep[see, e.g.,][]{Page2015}. 
We therefore make structural assumptions that imply that strata {\em (A)} through {\em (D)} do not exist, which reduces the number of possible strata from nine to five. 
First, we assume that there are no Defiers~\citep{angrist1996identification}; that is, there are no individuals who only enroll in ECHS if denied the opportunity to do so.
\begin{assumption}[No Defiers, or Monotonicity]\label{assump:no_defiers_binary}
There are no individuals with $\{ D_i(1) = lq, \ D_i(0) = e \}$ or $\{ D_i(1) = hq, \ D_i(0) = e \}$.
\end{assumption}
This eliminates strata {\em (A)} and {\em (B)}. 
To eliminate strata {\em (C)} and {\em (D)} we need an additional assumption:

\begin{assumption}[No Flip-Floppers]\label{assump:no_flip_floppers}
There are no individuals with $\{ D_i(1) = lq, \ D_i(0) = hq \}$ or $\{ D_i(1) = hq, \ D_i(0) = lq \}$.
\end{assumption}

This assumption states that individuals do not switch the type of non-ECHS school as a result of the ECHS lottery.~\citet{KlineWalters2014} refer to this as an independence of irrelevant alternatives assumption.
Applying Assumptions~\ref{assump:no_defiers_binary} and~\ref{assump:no_flip_floppers} leaves five remaining strata: ECHS Always Takers (\emph{eat}), Low-Quality Compliers (\emph{lc}), High-Quality Compliers (\emph{hc}), Low-Quality Always Takers (\emph{lat}), and High-Quality Always Takers (\emph{hat}), as shown in Table~\ref{tab:echs_table_5strata}. As we show in the supplementary materials, we can use these assumptions to identify the distribution of principal strata, $\pi_s$.
%
%

Next, we extend the standard exclusion restrictions~\citep[e.g.,][]{angrist1996identification} to the three ``Always'' strata in the more general setup:
\begin{assumption}[Exclusion restrictions]\label{assump:excl_rest}
There is no impact of randomization for individuals in the Always ECHS, Always Low-Quality, or Always High-Quality strata. That is,
$$\ITT_{eat} = \ITT_{lat} = \ITT_{hat} = 0.$$
\end{assumption}
The logic here is identical to the simpler noncompliance setting. That is, since randomization has no impact on school quality for students in these groups, we assume that randomization also has no impact on their later outcomes.
Finally, we can decompose the overall ITT effect into stratum-specific ITTs.
Under Assumptions~\ref{assump:no_defiers_binary},~\ref{assump:no_flip_floppers}, and~\ref{assump:excl_rest}:
\begin{align*} \text{Overall ITT }
        &= \pi_{lc}\ITT_{lc} + \pi_{hc} \ITT_{hc} + \pi_{eat} \ITT_{eat} + \pi_{lat} \ITT_{lat} + \pi_{hat} \ITT_{hat} \\
        &= \pi_{lc}\ITT_{lc} + \pi_{hc} \ITT_{hc}.  \numberthis \label{eq:overall_ITT}
\end{align*}
We can simplify this slightly by normalizing by the overall proportion of Compliers, $\pi_{lc} + \pi_{hc}$ :
\begin{align*}
\text{Overall LATE} &= \ITT_c \\
& = \frac{\pi_{lc}}{\pi_{lc} + \pi_{hc}} \ITT_{lc} + \frac{\pi_{hc}}{\pi_{lc} + \pi_{hc}} \ITT_{hc} \\
&= (1-\phi)~\ITT_{lc} + \phi~\ITT_{hc}, \numberthis \label{eq:overall_late}
\end{align*}
where $\phi = \frac{\pi_{hc}}{\pi_{lc} + \pi_{hc}}$ is the proportion of Compliers that have a High-Quality alternative.

We now have one equation and two unknowns. 
Without additional restrictions, we can only ``set identify'' the two impacts of interest, $\ITT_{lc}$ and $\ITT_{hc}$, as in~\citet{miratrix2017bounding}. 
In the next section, we discuss the use of cross-site variation to achieve point identification.
Other approaches are possible. 
First,~\citet{Feller2014} use a Bayesian model-based approach to estimate similar effects, though~\citet{Feller2014poster} suggest that such estimates might be unstable. 
Second,~\citet{mealli2016concentration} explore the use of multiple outcomes and other covariate restrictions.
Finally,~\citet{KlineWalters2014} identify these effects by imposing restrictions on the school type selection process.

\section{Identification and estimation via zero site-level correlation} \label{sec:identification_general}

We now turn to methods that exploit the multi-site experimental design to identify causal effects.
We introduce the core identifying assumption and the super-population of sites, and briefly discuss estimation, deferring many details to the supplementary materials. 

\subsection{Super-population of sites and the zero correlation assumption}\label{sec:site_super_pop}
We slightly extend our notation to emphasize the data's multi-site structure.
Let $k = 1,2,\dots,K$  index the $K$ sites of the experiment, where $X_i=k$ denotes that student $i$ belongs to experimental site $k$.
Let $\text{ITT}_{s|k} = \mathbb{E}\left[Y_i(1)-Y_i(0) | S_i=s, X_i=k \right ]$ be the impact of randomization for principal stratum $s$ in site $k$, with $\LATE_k = \ITT_{c|k}$;
let $\pi_{s|k} = \pr{S_i=s | X_i=k}$ be the proportion of individuals in principal stratum $s$ in site $k$;
and let $\phi_k = \pi_{hc|k}/(\pi_{lc|k} + \pi_{hc|k}$) denote the proportion of Compliers in site $k$ who are of High-Quality type. 
Our parameters of interest are the population average treatment impacts for Low-Quality Compliers and High-Quality Compliers, $\ITT_{lc}$ and $\ITT_{hc}$, for all students across all sites.

A key conceptual advance and statistical advantage of the multi-site setting, relative to a setting with a generic categorical covariate, is that we can envision a super-population of sites from which the $K$ observed sites are drawn. This is sometimes referred to as a random effects formulation~\citep[see, for example,][]{kolesar2015identification}, though we prefer to focus on the existence of a super-population.
Specifically, we assume that we sample sites represented as triples of parameters $\left( \ITT_{lc|k} , \; \ITT_{hc|k}, \; \phi_k \right)$ from an infinite super-population of sites with mean vector $\left(\ITT_{lc}, \ITT_{hc}, \phi \right)$ and a $3 \times 3$ correlation matrix $\bm \Sigma$:

\begin{equation}
\label{eq:super_pop}
\begin{pmatrix}
	\ITT_{lc|k} \\
	\ITT_{hc|k} \\
	\phi_k
\end{pmatrix} \stackrel{iid}\sim
\begin{bmatrix}
  \begin{pmatrix}
    \ITT_{lc} \\
    \ITT_{hc}  \\
    {\phi}
  \end{pmatrix}
 , \;\;
 \begin{pmatrix}
    \Sigma_{11} &  &  \\
    \Sigma_{21} & \Sigma_{22} &  \\
    \Sigma_{31} & \Sigma_{32} & \Sigma_{33} \\
  \end{pmatrix}
 \end{bmatrix}
\end{equation}

\noindent Under this interpretation, we extend the single super-population of individuals described in Section~\ref{sec:setup} to instead have two stages of sampling: first, we sample a site from an infinite super-population of sites; second, we sample an individual from the site-specific super-population.

Given this setup, it is natural to re-frame the main problem in terms of regression. 
First, re-write Equation~\eqref{eq:overall_late} separately for each site, re-arrange terms, and add zero twice to obtain
\begin{align*}
	\LATE_k \ \ = & \ \ (1-\phi_k)~\ITT_{lc|k} + \phi_k~\ITT_{hc|k} \\
	= & \ \ (1-\phi_k)~\ITT_{lc} + \phi_k~\ITT_{hc}~~+ \\
	& \qquad (1-\phi_k)~(\ITT_{lc|k} - \ITT_{lc}) + \phi_k~(\ITT_{hc|k} - \ITT_{hc}) \\
	= & \ \ (1-\phi_k)~\ITT_{lc} + \phi_k~\ITT_{hc} + (1-\phi_k) \ \epsilon_{lc|k} + \phi_k\ \epsilon_{hc|k}, \numberthis \label{eq:late_derive_randomness}
\end{align*}
\noindent where $\epsilon_{lc|k} = \ITT_{lc|k} - \ITT_{lc} $ and $\epsilon_{hc|k} = \ITT_{hc|k} - \ITT_{hc}$.
Across all $K$ sites, we therefore have a system of $K$ linear equations:
\begin{align*}
	\LATE_1 \ &=  \ (1-\phi_{1})\ITT_{lc} + \phi_{1}\ITT_{hc} + \eta_1  \\
	\LATE_2 \ &= \ (1-\phi_{2})\ITT_{lc} + \phi_{2}\ITT_{hc} + \eta_2 \\
	& \ \ \vdots \\
	\LATE_K \ &= \ (1-\phi_K)\ITT_{lc} + \phi_K\ITT_{hc} + \eta_K, \numberthis \label{eq:LATEols_zerocov}
\end{align*}
where we condense the final terms: $\eta_k = (1-\phi_k) \ \epsilon_{lc|k} + \phi_k \ \epsilon_{hc|k}$. 

This is a bivariate linear regression with no intercept, in which $\ITT_{lc}$ and $\ITT_{hc}$ are regression coefficients and $\eta_k$ is the regression error term. Since we have a super-population of sites, we can identify the causal effects of interest under the classical assumption that the regression errors, $\eta_k$ are uncorrelated with the regressors, $\phi_k$ and $1-\phi_k$, in the super-population. Specifically, we can identify the regression coefficients under the assumptions that $\Cov (\epsilon_{lc|k}, \phi_k) = 0$ and $\Cov (\epsilon_{hc|k}, \phi_k) = 0$, with the additional normalization that $\EE{\epsilon_{lc|k}} = 0$ and $\EE{\epsilon_{hc|k}} = 0$; or combining terms, $\Cov (\eta_k, \phi_k) = 0$ and $\EE{\eta_{k}}=0$.

%
%

\begin{assumption}[Zero site-level correlation between principal stratum distribution and principal causal effects]\label{assump:zero_corr_prop_LATE}
The site-specific relative share of High-Quality Compliers is uncorrelated with the site-specific impacts for High-Quality Compliers and for Low-Quality Compliers.
	\begin{equation}
		\Cov(\epsilon_{lc|k}~,~\phi_k
	) = 0
		\hspace{7pt}
		\text{and}
		\hspace{7pt}
		\Cov(\epsilon_{hc|k}~,~\phi_k
	) = 0 . \label{A:zero-corr-late}
	\end{equation}
\end{assumption}
\noindent This is equivalent to assuming that $\Sigma_{31} = \Sigma_{32} = 0$ in Equation~\eqref{eq:super_pop}.
In addition, we require that $\Var (\phi_k) > 0$, that is $\Sigma_{33} > 0$, which is analogous to the relevancy assumption in standard instrumental variables. 
We combine all these assumptions into the following proposition.

\begin{prop}[Identification of principal causal effects via zero site-level correlation]\label{prop:zero_corr_LATE}
For a multi-site trial with $K \geq 2$ sites, under assumption~\ref{assump:zero_corr_prop_LATE}, $\Var (\phi_k) > 0$, and the normalization that $\EE{ \epsilon_{lc|k} } = 0$ and $\EE{\epsilon_{hc|k} } = 0$, the principal causal effects, $\ITT_{lc}$ and $\ITT_{hc}$, are identified.
\end{prop}
\noindent The proof for Proposition~\ref{prop:zero_corr_LATE} follows immediately from standard regression theory.\footnote{These zero correlation and marginal zero expectation conditions are precisely the moment conditions needed to identify the regression coefficients in a linear regression model. A stronger assumption often cited for regression is \emph{strict exogeneity}, which states that the conditional mean of the error terms given the regressor equals zero,
$\E[\epsilon_{s|k} | \phi_k] = 0$. This assumption implies the two moment conditions above, but the reverse is not true; see~\citet{reardon2013_msmm_iv} for additional discussion in this context.} Importantly, while these results do not strictly require an underlying super-population of sites, it is difficult to imagine these conditions holding for a generic categorical covariate.

In the context of ECHS, the zero correlation assumption states that the impact of the program on High-Quality Compliers' ninth grade performance in a site does not systematically vary according to the relative proportion of High-Quality versus Low-Quality Compliers in a site; with the same assumption for Low-Quality Compliers. 
This strong assumption precludes factors that may differ across sites --- 
such as the average academic preparedness of incoming ninth grade students --- from influencing both the student compliance make-up of a site and the magnitude of impact ECHS has on students within the site.
Intuitively, students who are more academically prepared might have more resources and support, such that they would attend a High-Quality public school if they did not attend an ECHS.
In addition, students who enter ninth grade with a stronger academic background might experience ECHS differently from incoming students who have weaker academic foundations. 
To accommodate this kind of scenario, we discuss relaxing the zero-correlation assumption to hold conditional on covariates, such as prior academic preparedness, in Section~\ref{sec:aux_covariates}.

Finally, it is useful to re-frame this setup in terms of the contrast $\ITT_{hc} - \ITT_{lc}$. We can re-write Equation~\eqref{eq:LATEols_zerocov} to highlight this directly:
\begin{equation}
\LATE_k  =  \ITT_{lc} + \phi_k(\ITT_{hc} - \ITT_{lc})+ \eta_k, \mbox{ for } k = 1, \ldots, K. \label{eq:LATEols_zerocov_contrast}
\end{equation}
This yields a particularly simple form when there are only two sites, $j$ and $k$:
\begin{equation}
	\ITT_{hc} - \ITT_{lc} = \frac{\LATE_{j} - \LATE_k}{\phi_j - \phi_k}. \label{eq:ITThc-ITTlc_twosites}
\end{equation}
This is the slope of a line based on two points. It is also identical in form to the standard ratio estimator in instrumental variables, which underscores the connection to using the interaction of ``site by randomization'' as an additional instrument.
%
%
See the supplementary materials for additional discussion of restrictions with a binary covariate, including a discussion of the ASPES approach of~\citet{Peck2003}.


\subsection{Estimation}\label{sec:estimation}
In order to estimate these effects, we begin with an overly simplistic approach that uses plug-in estimators for the site-specific moments, $\widehat{\text{LATE}}_k$ and $\widehat{\phi}_k$. 
Let $\yhat_{zd} = \frac{1}{N_{zd}} \sum_{i \in \{ Z_i=z, \; D_i^{obs}=d \} } Y_i^{obs}$ be the finite sample average observed outcome for students assigned to $Z_i = z$ with observed take up $D_i^\obs = d$, 
and let $\yhat_{zd|k}$ be the corresponding estimate for students in site $k$. 
$\yhat_{z\cdot|k}$ indicates a summation over $d$; that is, the average observed outcome for students at site $k$ who were randomized to study arm $z$. 
Let $\widehat{\pi}_s$ denote the estimated proportion of individuals in principal stratum $s$, with $\widehat{\pi}_{s|k}$ the corresponding estimate for students in site $k$. 
(See the supplementary materials for details.)
We then estimate the site-specific LATE as
$$\widehat{\text{LATE}}_k = 
\frac{ \yhat_{1\cdot|k} - \yhat_{0\cdot|k} }{ \widehat{\pi}_{lc|k} + \widehat{\pi}_{hc|k} },$$
where $\widehat{\pi}_{lc|k} + \widehat{\pi}_{hc|k} $ is the estimated proportion of Compliers in site $k$. We can also estimate the relative proportion of High-Quality Compliers in site $k$:
$$\widehat{\phi}_k = \frac{ \widehat{\pi}_{hc|k} }{ \widehat{\pi}_{lc|k} + \widehat{\pi}_{hc|k} }.$$
With these site-aggregate statistics, we then estimate $\ITT_{lc}$ and $\ITT_{hc}$ via the regression coefficients from the site-level linear regression,
\begin{equation}
	\widehat{\text{LATE}}_k = \beta_{lc}~(1-\widehat{\phi}_k) + \beta_{hc}~\widehat{\phi}_k + \eta_{k} \ , \label{eq:ols-unadjusted}
\end{equation}
where $\widehat{\beta}_{lc}$ and $\widehat{\beta}_{hc}$ are estimators for $\ITT_{lc}$ and $\ITT_{hc}$, respectively.
Taking the site-specific estimates, $\widehat{\text{LATE}}_k$ and $\widehat{\phi}_k$, as fixed, we can account for uncertainty with the usual heteroskedastic-robust standard errors for linear regression~\citep{mackinnon1985}.

\paragraph{Measurement error.}
The plug-in approach ignores the fact that $\widehat{\text{LATE}}_k$ and $\widehat{\phi}_k$ are estimated rather than known. This leads to two key complications.
One complication is that conventional estimates of the standard error will under-estimate the true sampling variance. 
Also, the nominal point estimates could be biased; in particular, error in $\widehat{\phi}_k$ will attenuate the estimate of $\ITT_{hc} - \ITT_{lc}$. 
To account for the increased uncertainty due to measurement error, we therefore propose a straightforward case-resampling bootstrap approach that randomly samples students with replacement within each site.
For each bootstrap sample and independently for each site, we re-calculate $\widehat{\LATE}_k^\ast$ and $\widehat{\phi}_k^\ast$ and then estimate $\ITT_{lc}^\ast$ and $\ITT_{hc}^\ast$ via the linear model~\ref{eq:ols-unadjusted}. 
Finally, we apply standard multiple imputation combining rules~\citep{Rubin:1987} to obtain a single point estimate and standard error for each principal causal effect.

Extensive simulation studies (see supplementary materials) show that this procedure has meaningfully smaller RMSE than the naive procedure, but that bias in the point estimate is still problematic. 
Many alternatives are possible, such as a parametric bootstrap, which repeatedly draws $\widehat{\text{LATE}}_k^\ast$ and $\widehat{\phi}_k^\ast$ via a multivariate Normal with means and covariances estimated from each site. See the discussion in Section~\ref{sec:conclusion}.

\paragraph{Varying site size.} Finally, site sizes typically vary in practice, which introduces additional complications. Specifically, the super-population means $\left( \ITT_{lc}, \ITT_{hc}, \phi \right)$ discussed in Section~\ref{sec:site_super_pop} correspond to site-level averages.
If all sites have the same number of students, then the average over all sites equals the average over all students.
If site sizes vary, however, we must choose whether to weight sites equally (site average) or weight individuals equally (population average). 
Following~\citet{raudenbush2017}, when sites have different numbers of Compliers, the unweighted linear model~\ref{eq:ols-unadjusted} estimates the average principal causal effects across sites, rather than across individuals.
If, in addition to the conditions listed in Proposition~\ref{prop:zero_corr_LATE},
we also assert that $\ITT_{lc|k}$ and $\ITT_{hc|k}$ are independent of $N_k$, the number of Compliers in a site, then the population- and site-weighted estimates are equal. We return to this issue in the next section. 

\section{Conditional zero-correlation} \label{sec:aux_covariates}
In practice, we often observe a rich set of individual- and site-level covariates. 
While potentially helpful for increasing efficiency, such covariates are particularly useful for relaxing the unconditional zero correlation of Assumption~\ref{assump:zero_corr_prop_LATE}. 
Let ${\bf W}_k$ be a $w$-length vector of site-level covariates, which includes inherently site-level quantities, such as community type (urban, suburban, rural), as well as aggregate individual-level covariates, such as percent Free or Reduced-Price Lunch. 
We can then relax the zero correlation assumption such that it only holds {\em conditionally}:
\begin{align}
		\Cov{(\epsilon_{lc|k} \ , \ \phi_k\; | {\bf W}_k)} = 0 \;\;\; \text{and} \;\;\; \Cov{(\epsilon_{hc|k} \ , \ \phi_k\; | {\bf W}_k)} = 0 \ ,  \label{assump:zero_corr_condX_LATE}
\end{align}
with $\EE{\epsilon_{s|k} | {\bf W}_k}=0$, for $s \in \{lc, hc\}$. 
In the context of ECHS, this says, for example, that among sites of the same community type containing students of the same average level of academic preparedness, the impact of the ECHS program on different Complier types does not systematically vary according to the ratio of High- to Low-Quality Compliers in a site.
In general, to obtain consistent estimates for the principal causal effects, we want to condition on confounding factors of compliance and treatment impacts; that is, baseline covariates that are predictive of the distribution of principal strata in a site, and, separately, are predictive of the site-specific principal causal effects. 

There are several possible estimation procedures that incorporate auxiliary covariates under Assumption~\ref{assump:zero_corr_condX_LATE}.  
The most straightforward, given our regression setup, is to include (grand-mean centered) site-aggregate values of confounders as additional regressors in the site-level linear regression. Specifically, instead of fitting model~\ref{eq:ols-unadjusted}, we fit
\begin{equation}
	\widehat{\text{LATE}}_{k} =  \beta^{adj}_{lc}~(1-\widehat{\phi}_k) + \beta^{adj}_{hc}~\widehat{\phi}_k + {\bm \gamma}{\bf W}_k + \eta^{adj}_{k} \ . \label{eq:ols-adjusted}
\end{equation}
As above, ${\bf W}_k$ is a vector of site-aggregate covariate values, which could also include $N_k$, the total number of Compliers in site $k$. 
%

The simple regression-adjusted model, however, restricts the possible treatment effect variation; see supplementary materials for additional discussion. 
For example, if we believe a baseline covariate $W_{1,k}$ influences the impact of ECHS on student on-track status differently for a predominately High-Quality Complier site compared to a site with mostly Low-Quality Compliers, then we may prefer the interaction adjusted model
\begin{equation}
	\widehat{\text{LATE}}_{k} =  \beta^{int}_{lc}~(1-\widehat{\phi}_k)~ + ~\beta^{int}_{hc}~\widehat{\phi}_k ~ + ~{\bm \gamma}~{\bf W}_{-1,k}~ + ~{\delta_{lc}}(1-\widehat{\phi}_k) W_{1,k}~ + ~{\delta_{hc}}~\widehat{\phi}_k W_{1,k} ~ + ~ \eta^{int}_{k} \ , \label{eq:ols-int-adjusted}
\end{equation}
where appropriate combinations of $\widehat{\beta}^{int}_{s}$ and $\widehat{\delta}_s$ yield estimates of the site-average impacts.

Finally, when site sizes vary, we can re-weight the regression coefficient estimates from Eqs.~\eqref{eq:ols-adjusted} or \eqref{eq:ols-int-adjusted} to obtain population-average impacts under the assumption that $\ITT_{lc|k}$ and $\ITT_{hc|k}$ are \textit{conditionally} independent of $N_k$, the number of Compliers in a site, given ${\bf W}$. For High-Quality Compliers, we have the following weighted average:
%
\begin{equation}
	\widehat{\ITT}_{hc}^{pop} = \sum_{k=1}^K { \left( \widehat{\beta}^{int}_{hc} +  \widehat{\bm \gamma}~{\bf W}_{-1,k} + \widehat{\delta}_{hc}W_{1,k} \right) }  ~\frac{ \hat{\phi}_k N_k }{ \sum_{k=1}^K { \hat{\phi}_k N_k }}, \label{eq:ITThc_estimator}
\end{equation}
with an analogous estimate for Low-Quality Compliers.


\section{Analysis of ECHS} \label{sec:echs_results}

\subsection{Main analysis}
We investigate the impact of ECHS on the ninth grade on-track status of High-Quality Complier and Low-Quality Complier students.
As we discuss in Section~\ref{sec:site_super_pop}, we initially assume that the average impact of the Early College program on High-Quality Compliers' ninth grade performance is the same, in expectation, across all sites, and does not systematically vary according to the relative proportion of High-Quality versus Low-Quality Compliers in a site
(with the the same for Low-Quality Compliers).
We then relax this assumption by conditioning on standardized eighth grade reading score, which is predictive of both the relative proportion of High-Quality Compliers and of on-track percentages in sites.\footnote{Eighth grade reading score is also highly correlated with many of the other available covariates~\citep[see also][]{miratrix2017bounding}. Adjusting for all six available baseline covariates---student race, gender, free or reduced-price lunch eligibility, first generation college student status, and standardized eighth grade reading and math scores---yields meaningfully noisier estimates. An additional complication is that many of these lotteries are for the same ECHS program over multiple years. In principle, we could restrict the sample to schools with multiple lotteries and condition our analysis on the specific ECHS or specify a hierarchical model. In practice, this is infeasible with our limited number of sites.}

As described in the supplementary materials, we estimate impacts without covariate adjustment, with simple linear adjustment for site-average reading score, and with an interaction adjustment for site-average reading score. We account for different site sizes by taking weighted averages of predicted site-level impacts.

\begin{figure}[htb]
 \center
 \includegraphics[scale=0.5]{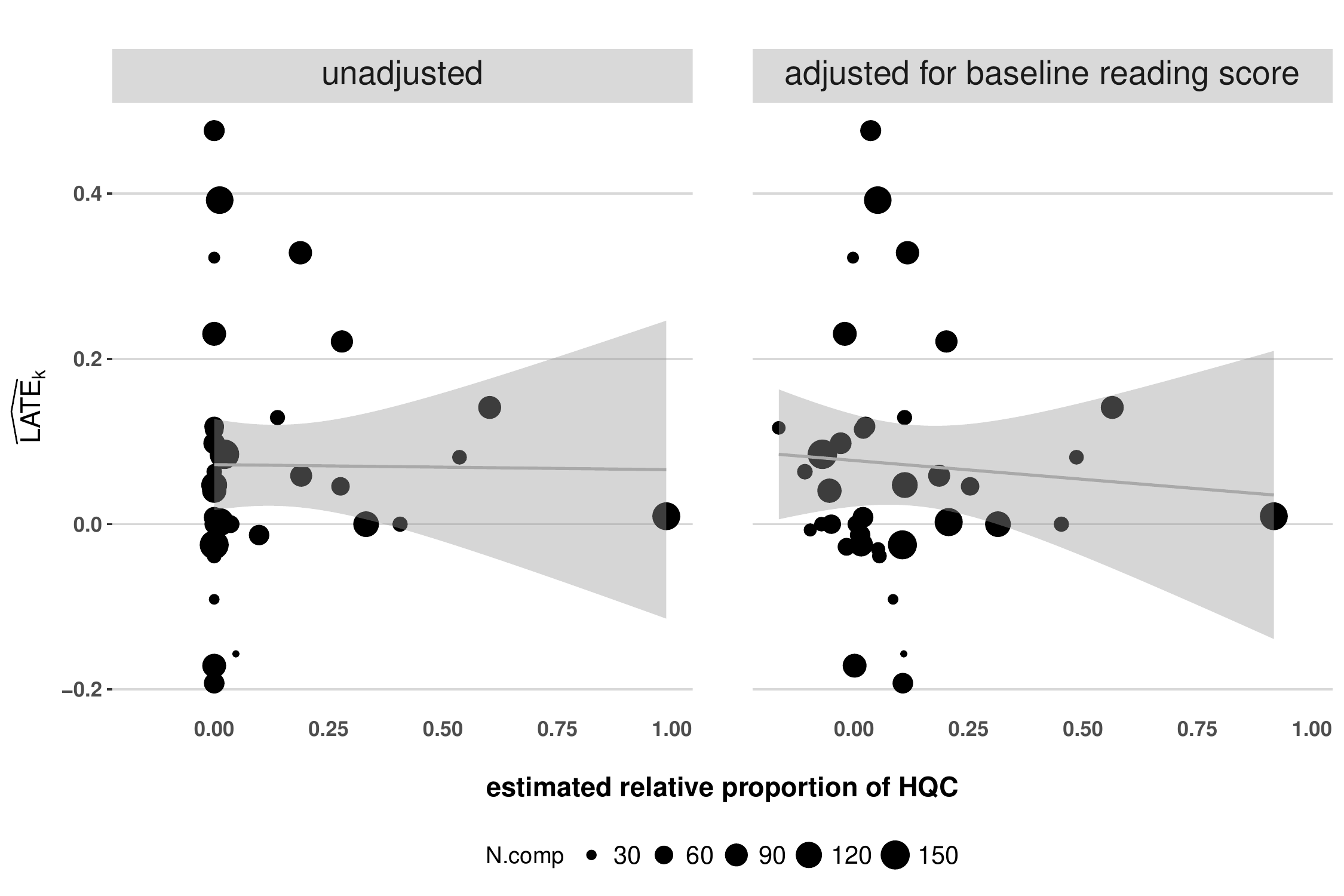}
  \caption{ {\bf ECHS site-level data.} Scatterplots of estimated site-specific Complier impacts (proportion on-track) versus (left panel) estimated relative proportion of High-Quality Compliers in a site, and (right panel) estimated \emph{residual} relative proportion of High-Quality Compliers in a site, after regressing $\hat{\phi}_k$ on eighth grade reading score. The size of the points indicate the number of Compliers in a site. The lines fit to the points correspond to linear regressions with a free intercept; the y-intercept for each line is an estimate for $\ITT_{lc}$, while the slope of each line is an estimate for the contrast $\ITT_{hc} - \ITT_{lc}$. The shaded grey regions are 95\% confidence intervals for the conditional mean outcome.}
   \label{fig:LATE_vs_phihat}
\end{figure}

\begin{figure}[htb]
 \centering
 \includegraphics[scale=0.45]{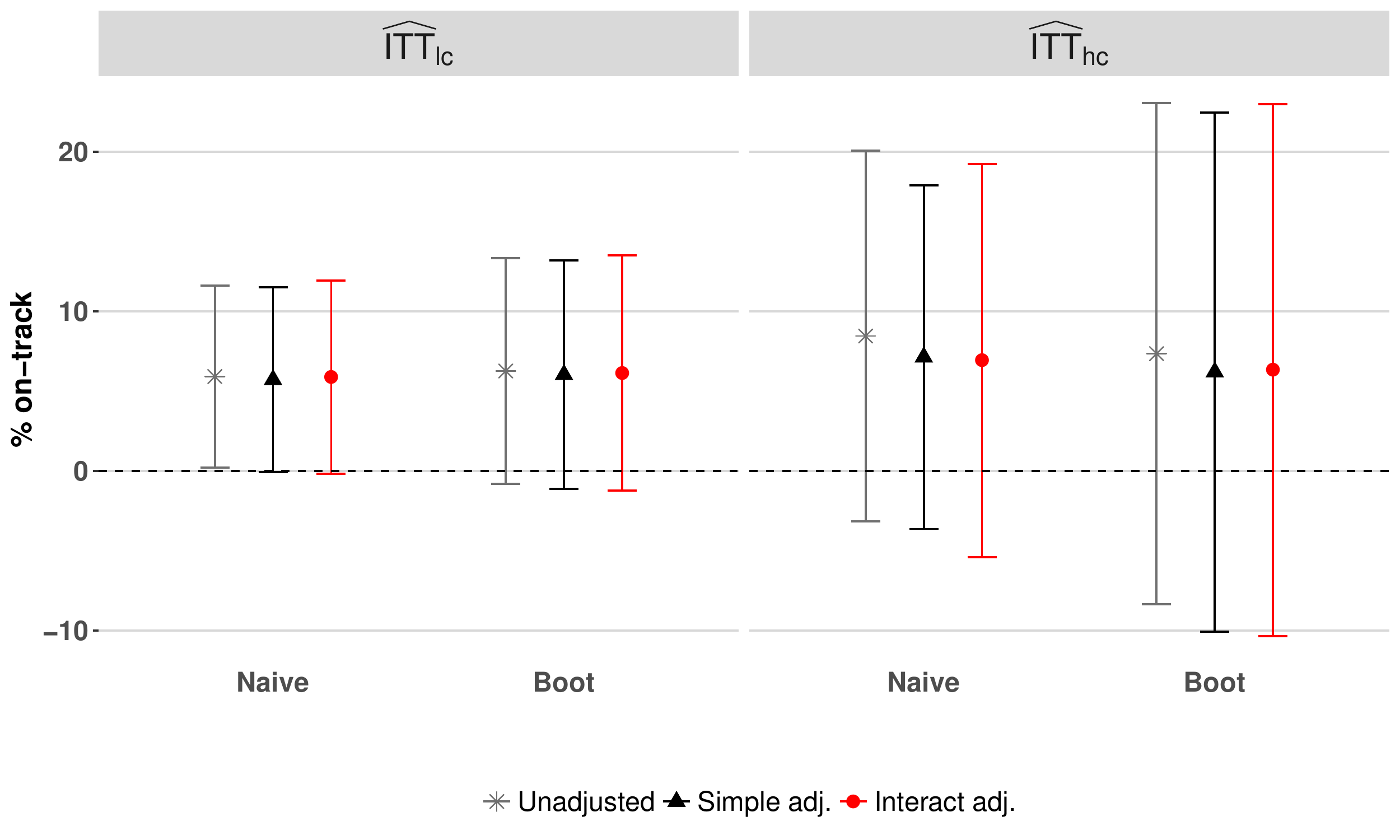}
  \caption{ {\bf Estimates of principal causal effects.} Point estimates and 95\% confidence intervals for Low- and High-Quality Complier principal causal effects are plotted for each estimation method.}
   \label{fig:echs_estimates_LATEmod}
\end{figure}

Figure~\ref{fig:LATE_vs_phihat} shows scatterplots of the estimated site-specific Complier impacts 
of ECHS on proportion on-track versus the estimated relative proportion of High-Quality Compliers in each site, 
before and after adjusting for site-average eighth grade reading score.
As the left panel shows, 22 of the 38 sites have an estimated $\widehat{\phi}_k = 0$, meaning that we estimate that all of the Compliers at these sites are Low-Quality Compliers. Since the Low-Quality Compliers are also the much larger group, we therefore anticipate more precise estimates of $\ITT_{lc}$ than $\ITT_{hc}$.
%
%

Figure~\ref{fig:echs_estimates_LATEmod} shows the corresponding point estimates and 95\% confidence intervals for $\ITT_{lc}$ and $\ITT_{hc}$.
All the point estimates are positive, between 5.7 and 8.5 percentage points.
There is no noticeable difference between the unadjusted versus simple adjusted or interaction adjusted point estimates for $\ITT_{lc}$; nor is there a meaningful difference between the naive and bootstrap point estimates.
Reading score adjustment has a more noticeable effect on point estimates for $\ITT_{hc}$, with $\widehat{\ITT}_{hc}$ decreasing by about 1.3 percentage points under both simple linear adjustment and interaction adjustment.

The standard errors for both $\widehat{\ITT}_{lc}$ and $\widehat{\ITT}_{hc}$ increase slightly under interaction adjustment, compared to no adjustment or simple adjustment.
For $\ITT_{lc}$ and $\ITT_{hc}$, respectively, the bootstrap CI for each adjustment method is roughly 23\% and 40\% wider than the CI of the corresponding naive estimate.
This aligns with our simulation study finding that the bootstrap method produces overly conservative confidence intervals.
Although we do not illustrate the results here, we note that adjusting for any single baseline covariate produces results that are substantively the same as those for reading score adjustment.
Finally, we assess whether there are meaningful differences between $\ITT_{hc}$ and $\ITT_{lc}$ using the re-parameterization in Equation~\eqref{eq:LATEols_zerocov_contrast}, which is illustrated by Figure~\ref{fig:LATE_vs_phihat}, in which the y-intercept is an estimate for $\ITT_{lc}$ and the slope for $\widehat{\phi}_k$ is an estimate for the difference $\ITT_{hc} - \ITT_{lc}$.
We do not find meaningful differences in stratum impacts for High- vs Low-Quality Compliers.

Overall, we find that the estimated impacts are quite similar for both Low- and High-Quality Compliers and that these estimates are stable across different models. Partly because the Low-Quality Complier group is larger, we are much more confident that the impact for this group is positive. By contrast, the estimated impact for High-Quality Compliers is much noisier. These results are consistent with the bounds in~\citet{miratrix2017bounding}.

\subsection{Model checking} 
An advantage of using a regression-based approach is that we can assess key identifying assumptions using standard regression diagnostics. 
In particular, the zero site-level correlation between principal stratum membership and stratum-specific impacts (Assumption~\ref{assump:zero_corr_prop_LATE}) 
implies that $\EE{\eta_{k}}=0$ and $\Cov({\eta_{k} , \phi_{k}}) = 0$. 
We can use the fitted residuals from the site-level regression to assess the evidence against these assumptions, though power might be limited.
Importantly, the zero-correlation assumption is restricted to \emph{mean} independence of the residual, rather than full stochastic independence. 
Thus, we would reject the identifying assumptions if there is a strong linear association, but would fail to reject even if there is, for example, meaningful evidence of heteroskedasticity. This approach is similar in spirit to tests for over-identifying restrictions in IV models~\citep[see, for example,][]{kolesar2015identification}. 

\begin{figure}[tb]
 \center
 \includegraphics[scale=0.5]{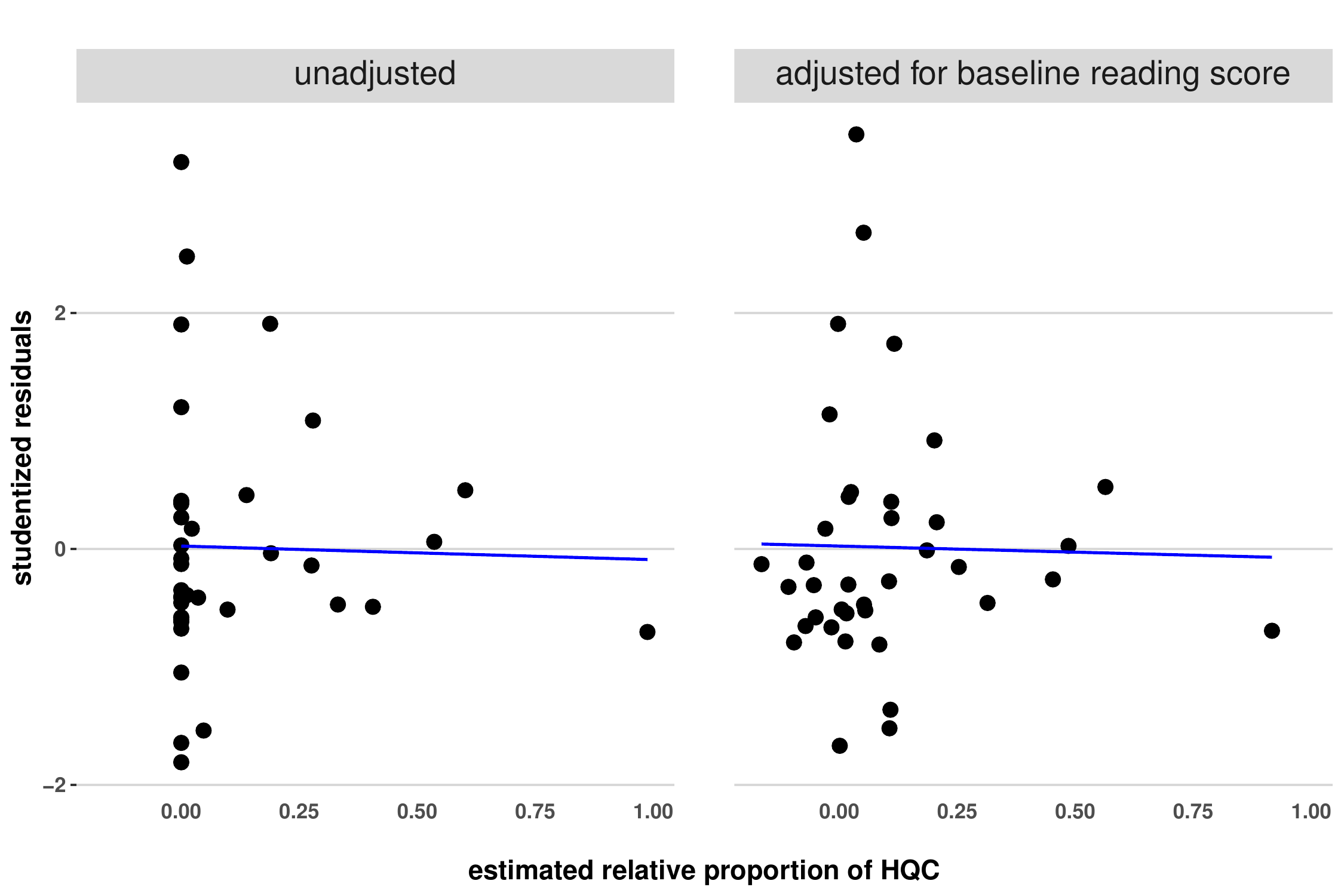}
  \caption{ {\bf Residual plots.} Studentized residuals versus estimated proportion of High-Quality Compliers for the Naive LATE model, where there is no baseline covariate adjustment (left panel) and where there is regression adjustment for eighth grade reading score (right panel). The blue lines are best-fit lines; one with a steep slope would indicate a violation of the (conditional) zero site-level correlation assumption needed to identify $\ITT_{lc}$ and $\ITT_{hc}$.}
   \label{fig:residuals}
\end{figure}

Figure~\ref{fig:residuals} shows studentized residual plots corresponding to the unadjusted and simple adjusted linear models (Equations~\ref{eq:ols-unadjusted} and~\ref{eq:ols-adjusted}) fit to the site-aggregate ECHS data shown in Figure~\ref{fig:LATE_vs_phihat}.
As indicated by the blue best-fit line for each residual plot, there is no strong positive or negative linear pattern to the residuals, and the means of the residuals for each model are close to zero.
Thus, there is no evidence against the identifying zero correlation assumptions, Assumptions~\ref{assump:zero_corr_prop_LATE} and~\ref{assump:zero_corr_condX_LATE}.
At the same time, the residual plots clearly invalidate a homogeneity assumption~\citep{Jiang2016} that the stratum-specific impacts are constant across sites,
with large changes in the conditional variance of the residuals across $\widehat{\phi}_{k}$.

\section{Connection to other methods} \label{sec:ties}
Several approaches have the same setup as what we explore here, but rest on stronger assumptions. 
First, we can impose a stronger version of Assumption~\ref{assump:zero_corr_prop_LATE} by assuming that average impacts are {\em constant} across sites, rather than equal \emph{in expectation} across sites.
Specifically, instead of assuming $\EE{\epsilon_{lc|k} \mid \phi_k} = 0$ for all $k$, we could instead require that $\epsilon_{lc|k} = 0$ for all $k$, or, equivalently, that $\ITT_{lc|1} = \cdots = \ITT_{lc|K}$. 
This clearly satisfies the requirements of Proposition~\ref{prop:zero_corr_LATE}, but is stronger than necessary for inference in our setting. 
Following the ecological inference literature, we refer to this as the \emph{constancy} assumption; 
\citet{gelman2001models} provides a discussion of the {\em constancy} versus {\em zero correlation} assumptions in ecological regression.
\citet{Jiang2016} instead call this constancy assumption the homogeneity assumption;  
\citet{wang2016identification} relax this assumption by adjusting for baseline covariates;~\citet{Kang2015} leverage this assumption to relax other requirements on possible effects. 

One conceptual advantage of this constancy assumption is that we no longer need to posit the existence of a (hypothetical) super-population of sites.
Instead, we can imagine sampling from an infinite super-population of individuals divided into $K$ fixed sites.
In fact, we no longer need multiple sites:
the assumption of constant impacts could be applied to a single-site experiment where we imagine sampling from an infinite super-population of individuals divided into $K$ fixed levels of any discrete covariate, such as grade level or racial group.
In practice, the estimators for $\ITT_{lc}$ and $\ITT_{hc}$ would be the same as in Section~\ref{sec:estimation}, even though the underlying assumption is much stronger. See, for example,~\citet{Hull2018}, who presents a similar setup as ours for a single site quasi-experiment with strata defined by a single (binary) covariate.\footnote{The core identifying assumption there is what Hull terms `LATE homogeneity', which says stratum-specific LATEs are mean independent of the stratifying covariate.}

The zero site-level correlation assumption we pose is also closely related to an important assumption in the multiple-site, multiple-mediator instrumental variables (MSMM-IV) literature.
For a multi-site study in which a treatment may affect the outcome through multiple mediators,~\citet{reardon2013_msmm_iv} delineate nine assumptions needed to identify the relevant causal effects using cross-site variation.
Of the nine assumptions, the authors emphasize the critical assumption of \emph{between-site compliance-effect independence}, in which the site-average compliance of each mediator is independent of the site Complier average effect of each mediator. 
This independence assumption is a closely related, but slightly stronger, version of the uncorrelatedness and marginal zero mean error conditions of Proposition~\ref{prop:zero_corr_LATE}.

Finally, we can re-frame much of the above discussion, such as Assumption~\ref{assump:zero_corr_prop_LATE}, in terms of site-level {\em means} rather than site-level impacts. 
That is, we could assume that the site-specific mean outcome of Low-Quality Compliers assigned to treatment is uncorrelated with the site-specific relative share of Low-Quality Compliers. 
We view this as a slightly stronger assumption than what we propose.
For example, it is conceivable that Low-Quality Complier students generally have less support and fewer resources that allow them to engage in academic activities, giving them a starting disadvantage compared to High-Quality Compliers. 
Thus, in schools with a larger share of Low-Quality Compliers, students' academic performance under no intervention could be poorer, on average, than student academic performance at schools composed mostly of High-Quality Compliers.
This scenario would violate a zero correlation in site-level means assumption. 
Assumption~\ref{assump:zero_corr_prop_LATE}, on the other hand, permits control mean outcomes to co-vary with the relative proportion of Low-Quality Compliers in a site.


\section{Conclusion} \label{sec:conclusion}

The principal stratification literature largely focuses on randomized studies where there is only one experimental site. We extend this framework to the multi-site setting in the context of an evaluation of Early College High Schools and show how to identify and estimate key principal causal effects under a strong zero correlation assumption. We relax this assumption by incorporating auxiliary covariates and explore several issues that arise in estimation. 

There are several directions for future work. The most important is to explore estimators that appropriately account for measurement error. 
First, we could adapt methods from the literature on multi-site, multi-mediator IV; specifically,~\citet{reardon2014bias} offer two bias-corrected instrumental variables estimators that could be extended to principal stratification. Second, we could further explore standard measurement error models or fully Bayesian hierarchical models as a way to simultaneously address both bias and sampling variance;~\citet{Bloom:2014va} discuss relevant strategies in the multi-site setting, including under noncompliance.

Finally, it is useful to assess how to incorporate the zero correlation assumption into a broader principal stratification analysis, such as a bounds approach~\citep{miratrix2017bounding}. Understanding the many possible identification and estimation approaches is increasingly important as more and more researchers use the principal stratification framework.

\clearpage

\bibliographystyle{imsart-nameyear}
\bibliography{covariates-for-PCE}


\newpage
\appendix
\appendixpage
 
\startcontents[sections]
\printcontents[sections]{l}{1}{\setcounter{tocdepth}{2}}

\clearpage

%
%

\section{Balance checks} \label{sec:balance_checks}

\begin{figure}[H]
 \centering
  \includegraphics[scale=0.5]{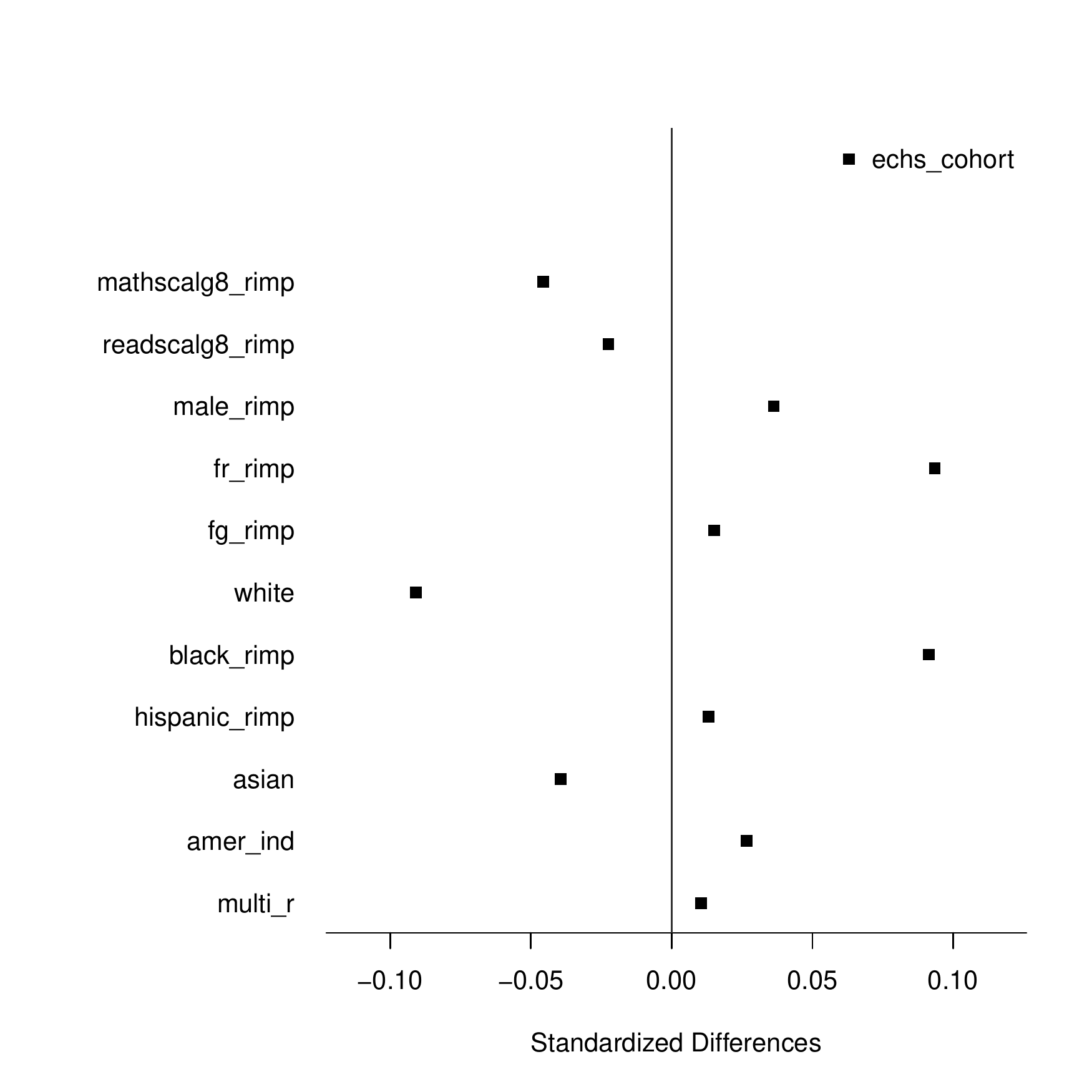}
 \caption{Standardized differences of treatment minus control, where group means are weighted by lottery size. There is slight imbalance in the free/reduced lunch price eligibility (fr\_rimp) and the white and black racial categories.}
 \label{fig:balance_checks}
\end{figure}

We perform baseline covariate balance checks for the ECHS data consisting of 3477 students ($N_t = 2021$, $N_c = 1456$) across 18 schools (each with up to 6 cohorts) and 38 lotteries.
Table~\ref{tab:balance_checks} shows the auxiliary covariates' lottery-size-weighted averages by treatment status.
The six baseline covariates are students' 8th grade math and reading scores (scaled, ranging from -4.17 to 3.53 for math, -3.95 to 2.88 for reading), and indicators for student gender, free or reduced-price lunch eligibility, first generation college student status, and race (a categorical variable with six levels). 
We detect a slight imbalance in the proportions of free/reduced-price lunch eligibility and the white and black racial categories across the two treatment arms.
We do not detect imbalance in any of the other baseline covariates (see Figure~\ref{fig:balance_checks}). 

\begin{table}[H]
\centering
\caption{Balance checks for ECHS baseline covariates}
\label{tab:balance_checks}
\begin{tabular}{lrrr}
\toprule
  & Treatment & Control & Std.diff\\
  \midrule
  math score & -0.053 & -0.007 & -0.046\\\\
  reading score & -0.015 & 0.007 & -0.022\\\\
  male & 0.414 & 0.396 & 0.036\\\\
  free/reduced-price lunch & 0.516 & 0.469 & 0.093\\\\
  first gen college student & 0.410 & 0.402 & 0.015\\\\
  \addlinespace
  white & 0.563 & 0.607 & -0.091\\\\
  black & 0.311 & 0.269 & 0.091\\\\
  hispanic & 0.078 & 0.075 & 0.013\\\\
  asian & 0.008 & 0.012 & -0.039\\\\
  american indian & 0.007 & 0.005 & 0.027\\\\
  multiracial & 0.033 & 0.031 & 0.011\\\\
  \bottomrule
  \end{tabular}
\end{table}

\subsection{Possible confounding factors}

\begin{figure}[H]
 \centering
  \includegraphics[scale=0.5]{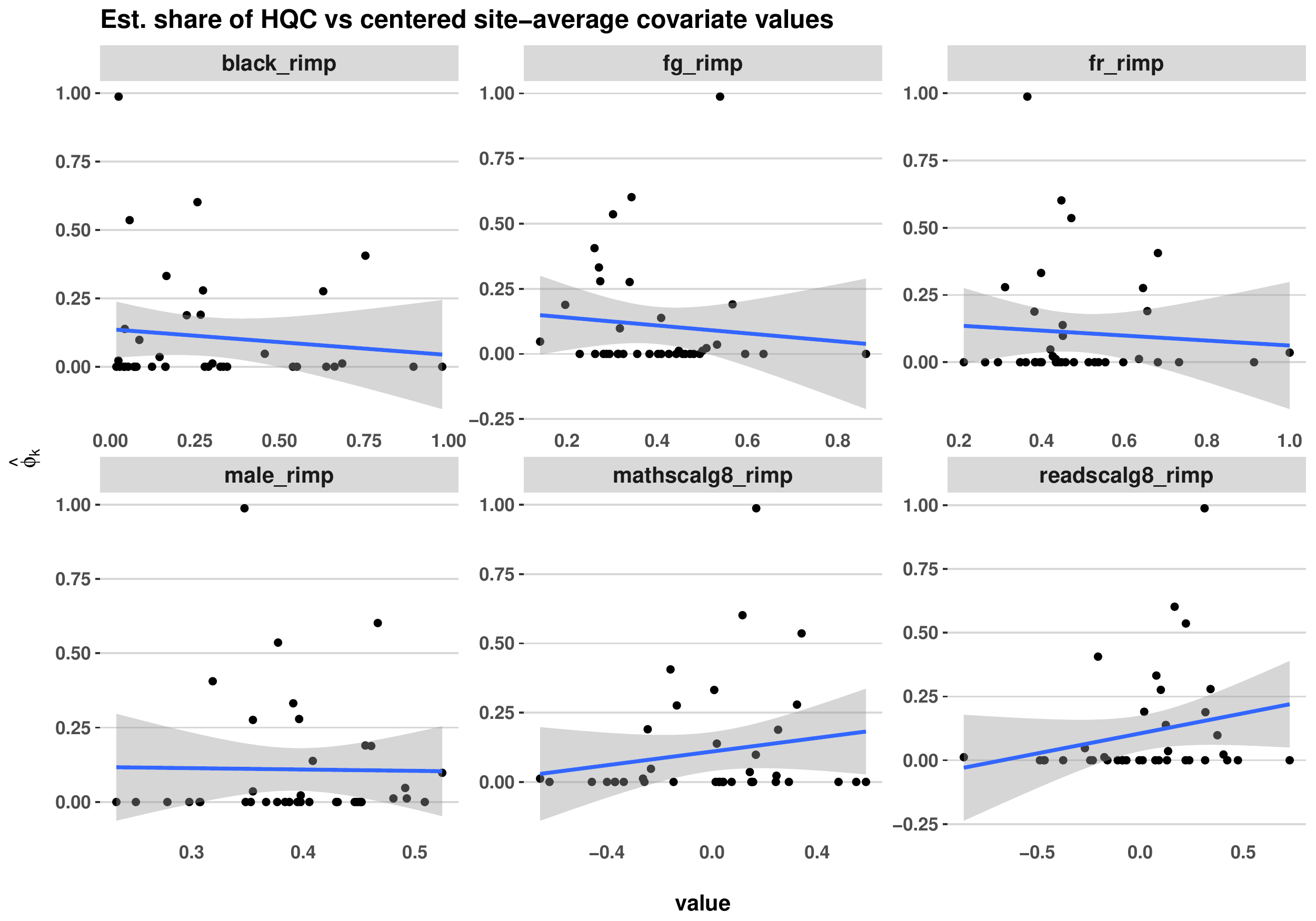}
 \caption{{\bf Linear regressions of estimated relative proportion of High-Quality Compliers against centered baseline covariate values}, across 38 lotteries. None of the baseline covariates are significantly predictive of $\widehat{\phi}_k$.}
 \label{fig:phihat_vs_X_supp}
\end{figure}

In our analyses, we consider adjusting for the six available baseline covariates mentioned above. 
To roughly quantify the plausibility of each covariate being a confounding factor of principal strata distribution and on-track rates, we regress estimated relative proportions of High-Quality Compliers on each centered baseline covariate, as illustrated in Figure~\ref{fig:phihat_vs_X_supp}.
In the dataset consisting of 38 lotteries, none of these $\hat{\phi}_k$ vs $X_k$ linear regressions indicate a significant correlation between the baseline covariate and relative proportion of High-Quality Compliers. 
In the full dataset of 44 lotteries, we find eighth grade math score and reading score to be predictive of the estimated share of High-Quality Compliers in a lottery.

Although we do not find these baseline covariates to be significantly linearly related to the estimated share of High-Quality Compliers, we nonetheless implement our estimation methods with and without covariate adjustment.
In alignment with existing recommendations in the covariate adjustment literature, our simulation study suggests that while adjusting for noisy covariates may decrease precision, adjustment should not lead to additional bias.
Failure to adjust for true confounding variables, however, will result in biased impact estimates.
Furthermore, adjusting for purely prognostic covariates will not increase bias, but may increase precision.

\section{Identifying and estimating principal strata in ECHS} \label{sec:echs_ps}

Under the general principal stratification setup in ECHS, we assume there are no Defiers or Flip-floppers, leaving us with five existing principal strata, $s \in \{ lc, \; hc, \; lat, \; hat, \; eat\}$, shown in Table~\ref{tab:echs_table_5strata}. 
Here, we describe how to identify and estimate these strata proportions, $\pi_s = \mathbb{P}[S_i = s]$, and individual stratum outcome means under treatment and control, $\mu_{s}(z)= \EE{Y_i(z) \mid S_i=s, Z_i = z}$.

\begin{table}[bt]
  \caption[Diagram of assumed principal strata in ECHS study]{The nine possible principal strata in the ECHS study. We assume that strata {\em (A)} - {\em(D)} do not exist, leaving five principal strata. The two highlighted cells indicate the strata of interest.}
   \label{tab:echs_table_5strata_supp}
	\begin{tabular}{c|M{20mm}|M{27mm}|M{27mm}|M{27mm}|N}
		\multicolumn{2}{c}{} & \multicolumn{3}{c}{\cen \ \ \ \bf No ECHS offer ($Z_i=0$)} &
		\\[10pt] \cline{3-5}
		\multicolumn{1}{c}{} & & \multicolumn{1}{c|}{$D_i(0)=e$} & \multicolumn{1}{c|}{$D_i(0)=lq$} & \multicolumn{1}{c|}{$D_i(0)=hq$} &
		\\[20pt] \cline{2-5}
		\multirow{4}{*}{\rotatebox[origin=c]{0}{\parbox[c]{1.5cm}{\cen \bf ECHS offer ($Z_i=1$)}} } & $D_i(1)=e$ & \textbf{ECHS Always Taker} & \cellcolor{lightgray} \textbf{Low-Quality Complier} & \cellcolor{lightgray} \textbf{High-Quality Complier} &
		\\[20pt] \cline{2-5}
		& $D_i(1)=lq$ & {\em (A)} & \textbf{Low-Quality Always Taker} & {\em (C)} &
		\\[20pt] \cline{2-5}
		& $D_i(1)=hq$ &{\em (B)} & {\em (D)} & \textbf{High-Quality Always Taker} &
		\\[20pt] \cline{2-5}
	\end{tabular}
\end{table}

\subsection{Identifying strata proportions in terms of directly estimable quantities} \label{sec:strata_proportions}

Ideally, we would like to observe the principal stratum category of each student, $S_i$. This is impossible, since we cannot observe both $D_i(0)$ and $D_i(1)$ for each student. 
We do, however, have $D_i^{obs} = D_i(Z_i)$, the observed school type student $i$ attends under treatment assignment $Z_i=z$, and $Y_i^{obs} = Y_i(Z_i)$, student $i$'s observed on-track status at the end of ninth grade.
We observe a total of six groups of students, defined by combinations of treatment assignment and subsequent school type attendance, which we index with $zs$, $z_i \in \{0, 1\}$ and $s_i \in \{e, hq, lq\}$.
Each of these six groups are mixtures of our five principal strata of interest, but due to random treatment assignment and SUTVA, we can use observed finite averages to separate out these mixtures and calculate quantities that are subsequently used to estimate our principal strata proportions and stratum treatment and control mean outcomes.

First, we identify the expected observed subgroup outcomes of all students who, if assigned to treatment $z$, would go to school-type $d$: 

\[
	\ybar_{zd} = \EE{Y_i^{obs}\; | \; Z_i=z, \; D_i^{obs}=d },
\]
which we can directly estimate using the finite sample average observed outcomes $\yhat_{zd} = \frac{1}{N_{zd}}\sum_{i \in \{Z_i=z, \; D_i^{obs}=d\} } Y_i^{obs}$, where $N_{zd} = \sum_{i=1}^N \mathbb{1}\{D_i^\obs(z) = d \}$ is the observed number of students who enroll in school-type $d$ when assigned treatment $z$.
Second, we identify the proportion of all students who would enroll in school-type $d$ if assigned treatment $z$: 

\[
	p_{zd} = \mathbb{P}[D_i(z) = d \mid Z_i = z],
\]
which we can directly estimate as $\hat{p}_{zd} = N_{zd}/N$. 
Note that $\hat{p}_{ze} + \hat{p}_{zlq} + \hat{p}_{zhq} = 1$ for $z = 0, 1$. 
Given identification of $p_{zd}$, we can proceed to identify the principal stratum proportions $\pi_{s}$, for $s \in \{ lc, \; hc, \; lat, \; hat, \; eat\}$. 
The identifying equations are displayed in Table~\ref{tab:echs_table_proportions_supp}. 
The three Always Takers strata are immediately identified from observed groups of student proportions; due to random assignment and SUTVA, for each of the student-types, we expect the proportion of that student-type to be the same across treatment and control groups. 
Thus, for example, the observed proportion of students who are assigned to control but still enroll in an ECHS ($\hat{p}_{0e}$) is equal in expectation to the proportion of ECHS Always Takers in the whole student population, $\pi_{eat}$. 
Similarly, the observed proportion of students who are given an ECHS offer but decide to enroll in an alternative Low-Quality public school is equal in expectation to the proportion of Low-Quality Always Takers in the superpopulation.
Among Compliers, those students who enroll in a Low-Quality public school under control are either Low-Quality Compliers or Low-Quality Always Takers. 
Analogously, students who enroll in a High-Quality public school under control are either High-Quality Compliers or High-Quality Always Takers. 
This leads to the equalities $p_{0lq} = \pi_{lc} + \pi_{lat}$ and $p_{0hq} = \pi_{hc} + \pi_{hat}$. 
Plugging in estimates, we thus have 

\begin{align*}
	\hat{\pi}_{lc} &= \hat{p}_{0lq} -\hat{\pi}_{lat} = \hat{p}_{0lq} - \hat{p}_{1lq} = \left( N_{0lq} - N_{1lq} \right) / N \\
	\hat{\pi}_{hc} &= \hat{p}_{0hq} - \hat{\pi}_{hat} = \hat{p}_{0hq} - \hat{p}_{1hq} = \left(N_{0hq} - N_{1hq} \right) / N.
\end{align*}

Analogous calculations can be performed within each site to estimate site-specific stratum proportions $\pi_{s|k}$.

\begin{table}[bt]
  \caption[Principal strata proportions in ECHS study]{Relationship between principal strata proportions, $\pi_s$, and proportions of students who, if assigned treatment $z$, would attend school-type $d$, $p_{zd}$. The two highlighted cells indicate the strata of interest.}
   \label{tab:echs_table_proportions_supp}
	\begin{tabular}{c|M{20mm}|M{28mm}|M{28mm}|M{28mm}|N}
		\multicolumn{2}{c}{} & \multicolumn{3}{c}{\cen \ \ \ \bf No ECHS offer ($Z_i=0$)} &
		\\[10pt] \cline{3-5}
		\multicolumn{1}{c}{} & & \multicolumn{1}{c|}{$D_i(0)=e$} & \multicolumn{1}{c|}{$D_i(0)=lq$} & \multicolumn{1}{c|}{$D_i(0)=hq$} &
		\\[20pt] \cline{2-5}
		\multirow{4}{*}{\rotatebox[origin=c]{0}{\parbox[c]{1.5cm}{\cen \bf ECHS offer ($Z_i=1$)}} } & $D_i(1)=e$ & $\pi_{eat} = p_{0e}$ & \cellcolor{lightgray} $\pi_{lc} = p_{0lq} - p_{1lq}$ & \cellcolor{lightgray} $\pi_{hc} = p_{0hq} - p_{1hq}$ &
		\\[20pt] \cline{2-5}
		& $D_i(1)=lq$ & {\em (A)} & $\pi_{lat} = p_{1lq}$ & {\em (C)} &
		\\[20pt] \cline{2-5}
		& $D_i(1)=hq$ & {\em (B)} & {\em (D)} & $\pi_{hat} = p_{1hq}$ &
		\\[20pt] \cline{2-5}
	\end{tabular}
\end{table}


\section{LATE vs ITT model} \label{sec:LATE_vs_ITT} 
Throughout our paper, we motivate our estimation approach using the decomposition of the overall LATE (Eq. 3.2 in the main text).
We can motivate similar approaches, however, using the decomposition of overall ITT (Eq. 3.1 in the main text).

Concretely, we can motivate our linear regression estimation framework using either of the following identities: 
\begin{align*} 
\text{Overall ITT }
  &= \pi_{lc} \text{ITT}_{lc} + \pi_{hc}\text{ITT}_{hc} +  \pi_{eat} \text{ITT}_{eat} +  \pi_{lat}   \text{ITT}_{lat} + \pi_{hat} \text{ITT}_{hat} \\  \numberthis \label{eq:overall_ITT_supp}
  &= \pi_{lc} \text{ITT}_{lc} + \pi_{hc}\text{ITT}_{hc} \ , 
\end{align*}
or its normalized version:
\begin{align*}
  \text{Overall LATE} &= \text{ITT}_c \\  \numberthis \label{eq:overall_LATE_supp}
  & = \frac{\pi_{lc}}{\pi_{lc} + \pi_{hc}} \text{ITT}_{lc} + \frac{\pi_{hc}}{\pi_{lc} + \pi_{hc}}       \text{ITT}_{hc} \\   \nonumber
  &= (1-\phi)~\text{ITT}_{lc} + \phi~\text{ITT}_{hc} \ , 
\end{align*}
where $\phi = \frac{\pi_{hc}}{\pi_{lc} + \pi_{hc}}$ is the proportion of Compliers who are High-Quality.
Either identity will permit us to estimate $\text{ITT}_{lc}$ and $\text{ITT}_{hc}$ using an ordinary least squares model.

Using the Overall ITT identity of Eq.~\eqref{eq:overall_ITT_supp}, we have for each site $k$,
\begin{align*}
	\ITT_k \ \ = & \ \ \pi_{lc|k}~\ITT_{lc|k} + \pi_{hc|k}~\ITT_{hc|k} \\
	= & \ \ \pi_{lc|k}~\ITT_{lc} + \pi_{hc|k}~\ITT_{hc}~~+ \\
	& \qquad \pi_{lc|k}~(\ITT_{lc|k} - \ITT_{lc}) + \pi_{hc|k}~(\ITT_{hc|k} - \ITT_{hc}) \\
	= & \ \ \pi_{lc|k}~\ITT_{lc} + \pi_{hc|k}~\ITT_{hc} + \pi_{lc|k} \ \epsilon_{lc|k} + \pi_{hc|k}\ \epsilon_{hc|k} \\
	= & \ \ \pi_{lc|k}~\ITT_{lc} + \pi_{hc|k}~\ITT_{hc} + \psi_{k} \numberthis \label{eq:ITT_derive_randomness_supp}
\end{align*}
\noindent where $\epsilon_{lc|k} = \ITT_{lc|k} - \ITT_{lc} $ and $\epsilon_{hc|k} = \ITT_{hc|k} - \ITT_{hc}$, and $\psi_{k} = \pi_{lc|k} \ \epsilon_{lc|k} + \pi_{hc|k}\ \epsilon_{hc|k}$.

Next, we invoke a slightly modified version of Assumption 4.1 from the main text, 
where instead of assuming that the site-specific \emph{relative share} of High-Quality Compliers is uncorrelated with the site-specific principal causal effects, we assume the uncorrelatedness holds between the \emph{absolute proportions} of High-Quality and Low-Quality Compliers and the site-specific principal impacts. 
Namely, we posit that $\EE{ \epsilon_{lc|k} } = 0$ and $\EE{\epsilon_{hc|k} } = 0$, and
\begin{equation}
		\Cov(\epsilon_{lc|k}~,~\pi_{lc|k}
	) = 0
		\hspace{7pt}
		\text{and}
		\hspace{7pt}
		\Cov(\epsilon_{hc|k}~,~\pi_{hc|k}
	) = 0 . \label{A:zero-corr-ITT_supp}
\end{equation}

Applying these assumptions to~\eqref{eq:ITT_derive_randomness_supp}, we can then fit the zero-intercept OLS model
\begin{equation}
\widehat{\text{ITT}}_k = \beta_{lc}\hat{\pi}_{lc|k} + \beta_{hc}\hat{\pi}_{hc|k} + \psi_k \ .  \label{eq:ITT_model_supp}
\end{equation}
Regression coefficients $\beta_{lc}$ and $\beta_{hc}$ would be estimators for $\ITT_{lc}$ and $\ITT_{hc}$, respectively.
The main text describes how to estimate the principal causal effects under the Overall LATE formulation of Eq.~\eqref{eq:overall_LATE_supp}.

While the Overall ITT and Overall LATE identities are equivalent, in practice the treatment impact estimates one calculates using model~\ref{eq:ITT_model_supp} versus its $\widehat{\LATE}_k$ cousin will not be exactly the same.
However, our simulation studies showed no meaningful difference between estimates obtained from the LATE vs ITT models. 
We chose to highlight the LATE version of the OLS model 
because it can be conveniently reparameterized to target contrasts of $\ITT_{lc}$ and $\ITT_{hc}$.

\subsection{Fitting LATE version of OLS model, under three parameterizations}  \label{sec:LATEmod_3versions}

The decomposition of overall LATE consists of two principal stratum weights that sum to one (see model~\ref{eq:overall_LATE_supp}).
This allows for three alternative parameterizations of the identity, each of which imply an OLS model with regression coefficients that serve as estimators for the principal causal effects.

The parameterization of~\eqref{eq:overall_LATE_supp} immediately implies a zero-intercept OLS model, as in
\begin{equation}
	\widehat{\text{LATE}}_k = \beta_{lc}(1-\hat{\phi}_k) + \beta_{hc}\hat{\phi}_k + \eta_k \ , \label{eq:ols-unadjusted_supp}
\end{equation}
where $(1-\widehat{\phi}_k)$ is coded as a separate variable from $\hat{\phi}_k$.
Then, the regression coefficient for $(1-\hat{\phi}_k)$ is an estimator for $\text{ITT}_{lc}$; the regression coefficient for $\hat{\phi}_k$ is an estimator for $\text{ITT}_{hc}$.

A second parameterization of \eqref{eq:overall_LATE_supp} is
\begin{equation}
  \text{Overall LATE} = \text{ITT}_{lc} + \phi (\text{ITT}_{hc} - \text{ITT}_{lc}).  \label{eq:overall_late_v2}
\end{equation}
This corresponds to an unconstrained linear regression 
\begin{equation}
	\widehat{\LATE}_k = \beta_{lc} + \beta_{h-l}\hat{\phi}_k + \eta_k . \label{eq:LATEmod_phi} \
\end{equation}
The intercept of model~\ref{eq:LATEmod_phi} is an estimator for $\text{ITT}_{lc}$ while the regression coefficient for $\hat{\phi}_k$ is an estimator for the contrast $\text{ITT}_{hc} - \text{ITT}_{lc}$. 

A third parameterization of Eq.~\eqref{eq:overall_LATE_supp} is

\begin{equation}
  \text{Overall LATE} = \text{ITT}_{hc} + (1-\phi) (\text{ITT}_{lc} - \text{ITT}_{hc}).  \label{eq:overall_late_v3}
\end{equation}
This equation corresponds to the fitted model
\begin{equation}
	\widehat{\LATE}_k  = \beta_{hc} + \beta_{l-h}(1-\hat{\phi}_k) + \eta_k. \label{eq:LATEmod_phic}  \
\end{equation}
From~\eqref{eq:LATEmod_phic}, the intercept is an estimator for $\text{ITT}_{hc}$ while the regression coefficient for $(1 - \hat{\phi}_k)$ is an estimator for the contrast $\text{ITT}_{lc} - \text{ITT}_{hc}$.

Thus, to readily obtain the point estimate and standard error of $\text{ITT}_{lc}$ and $\text{ITT}_{hc}$, we can fit one zero-intercept OLS model~\ref{eq:ols-unadjusted_supp} and use the regression coefficients from that model as estimators for our estimands of interest. 
Alternatively, we can fit two unconstrained OLS models (\ref{eq:LATEmod_phi} and \ref{eq:LATEmod_phic}) and take the intercepts from those two models as estimators for $\text{ITT}_{lc}$ and $\text{ITT}_{hc}$. 
The point estimates and standard errors under these two approaches are equivalent. Note that this equivalence holds even when we adjust for other pretreatment covariates.

\section{Incorporating auxiliary covariates}\label{sec:aux_covariates_supp}

\subsection{Identification}
In practice, we often observe a rich set of individual- and site-level covariates. 
While potentially helpful for increasing efficiency, such covariates are especially useful for relaxing the unconditional zero correlation of Assumption 4.1.
In particular, let ${\bf W}_k$ be a $w$-length vector of site-level covariates, which includes inherently site-level quantities, such as community type (urban, suburban, rural), as well as aggregate individual-level covariates, such as percent Free or Reduced-Price Lunch. 
We can relax the zero correlation assumption such that it only holds {\em conditionally}:
\begin{align}
		\Cov{(\epsilon_{lc|k} \ , \ \phi_k\; | {\bf W}_k)} = 0 \;\;\; \text{and} \;\;\; \Cov{(\epsilon_{hc|k} \ , \ \phi_k\; | {\bf W}_k)} = 0 \ ,  \label{assump:zero_corr_condX_LATE_supp}
\end{align}
with $\EE{\epsilon_{s|k} | {\bf W}_k}=0$, for $s \in \{lc, hc\}$. 
In the context of ECHS, this says, for example, that among sites of the same community type containing students of the same average level of academic preparedness, the impact of the ECHS program on different Complier types does not systematically vary according to the ratio of High- to Low-Quality Compliers in a site.
In general, to obtain consistent estimates for the principal causal effects, we want to condition on confounding factors of compliance and treatment impacts; that is, baseline covariates that are predictive of the distribution of principal strata in a site, and, separately, are predictive of the site-specific principal causal effects.

\subsection{Estimation} \label{sec:auxiliaryXs_supp}
There are several possible approaches to adjust for confounding variables in our setting. 
The most straightforward, given our regression setup, is to include (grand-mean centered) site-aggregate values of confounders as additional regressors in the site-level linear regression. 
Specifically, instead of fitting model~\ref{eq:ols-unadjusted_supp}, we fit

\begin{equation}
	\widehat{\text{LATE}}_{k} =  \beta^{adj}_{lc}~(1-\widehat{\phi}_k) + \beta^{adj}_{hc}~\widehat{\phi}_k + {\bm \gamma}{\bf W}_k + \eta^{adj}_{k} \ . \label{eq:ols-adjusted_supp}
\end{equation}
The superscript ${adj}$ reminds us that the regression coefficients are from a model that does simple covariate adjustment.
As above, ${\bf W}_k$ is a vector of site-aggregate covariate values.
If $N_k$, the total number of Compliers in site k, is correlated with the site-specific distribution of principal strata, then we can adjust for this by including $N_k$, or some transformation of $N_k$, as a regressor in model~\ref{eq:ols-adjusted_supp}.

Estimates of the population average treatment impacts for High- and Low-Quality Compliers are then sample size weighted means of predicted site-level impacts, but with a design matrix that picks out the regression coefficients corresponding to Low- or High-Quality Compliers:
\begin{equation}
	\widehat{\ITT}_{lc} =  {\bm w}^{lc}~\mathbb{1}_{(1 \times K)}~{\bm X}_{lc}~\widehat{\bm \beta};\label{eq:popu_ITTlc_withX_supp}
\end{equation}

\begin{equation}
	\widehat{\ITT}_{hc} =  {\bm w}^{hc}~\mathbb{1}_{(1 \times K)}~{\bm X}_{hc}~\widehat{\bm \beta}, \label{eq:popu_ITThc_withX_supp}
\end{equation}
where ${\bm X_{lc}} = \left( \bm 1~ \bm 0~ \bm 1_{(K \times w)} \right)$,
${\bm X_{hc}} = \left( \bm 0~ \bm 1~ \bm 1_{(K \times w)} \right)$, and 
$\widehat{\bm \beta} = \left(\widehat{\beta}^{adj}_{lc},~ \widehat{\beta}^{adj}_{hc},~\widehat{ \bm \gamma} \right)^\top$ is the $(2+w)$-length vector of regression coefficients from model~\ref{eq:ols-adjusted_supp}.

Depending on the strength and direction of the confounding variable(s) and their functional form with respect to compliance and treatment impact, it may be necessary to include interaction terms of $\widehat{\phi}_k$ and ${\bf W}_k$, and/or quadratic or higher-order terms in the regression.
For example, if we believe a baseline covariate $W_{1,k}$ influences the impact of ECHS on student on-track status differently for a predominately High-Quality Complier site compared to a site with mostly Low-Quality Compliers, then we may prefer the interaction adjusted model
\begin{equation}
	\widehat{\text{LATE}}_{k} =  \beta^{int}_{lc}~(1-\widehat{\phi}_k)~ + ~\beta^{int}_{hc}~\widehat{\phi}_k ~ + ~{\bm \gamma}~{\bf W}_{-1,k}~ + ~{\delta_{lc}}(1-\widehat{\phi}_k) W_{1,k}~ + ~{\delta_{hc}}~\widehat{\phi}_k W_{1,k} ~ + ~ \eta^{int}_{k} \ . \label{eq:ols-int-adjusted_supp}
\end{equation}
Writing out the matrix calculations of Eqs.~\eqref{eq:popu_ITTlc_withX_supp} and~\eqref{eq:popu_ITThc_withX_supp} as summations, our estimate for the population average treatment impact for Low-Quality Compliers is thus
\begin{equation}
	\widehat{\ITT}_{lc} = \sum_{k=1}^K { \left( \widehat{\beta}^{int}_{lc} +  \widehat{\bm \gamma}~{\bf W}_{-1,k} + \widehat{\delta}_{lc}W_{1,k} \right) }  ~\frac{ (1-\hat{\phi}_k)N_k }{ \sum_{k=1}^K { (1-\hat{\phi}_k)N_k }}. \label{eq:ITTlc_estimator_supp}
\end{equation}
Analogously, for High-Quality Compliers,
\begin{equation}
	\widehat{\ITT}_{hc} = \sum_{k=1}^K { \left( \widehat{\beta}^{int}_{hc} +  \widehat{\bm \gamma}~{\bf W}_{-1,k} + \widehat{\delta}_{hc}W_{1,k} \right) }  ~\frac{ \hat{\phi}_k N_k }{ \sum_{k=1}^K { \hat{\phi}_k N_k }}. \label{eq:ITThc_estimator_supp}
\end{equation}

Standard errors for these sample size weighted estimators are calculated by taking design-matrix-weighted sums of the elements of the heteroskedastic-robust~\citep{mackinnon1985} covariance matrix from the fitted linear regression.

Finally, we explored using pair matching rather than regression to adjust for ${\bf W}$; see, for example,~\citet{zubizarreta2013}. However, we did not find this approach to yield useful results in our settings of interest.

%
%

\section{Connection to Analysis of Symmetrically Predicted Endogenous Subgroups (ASPES)}

For the setting of a single-site randomized-controlled experiment,~\citet{Peck2003} and~\citet{Peck:2013jt} propose ASPES, a two-stage procedure for measuring program impacts on subgroups identified by post-treatment traits.~\citet{Bein:2015ud} connects this approach to principal stratification.

The first stage involves estimating a model for compliance status using baseline features of a random subset of the treatment group, for whom we observe a certain type of program participation such that we know their compliance membership.
This fitted compliance model, or \emph{principal score} model~\citep{Feller2016b}, is then used to predict compliance status for the remaining treatment group members and also for the control group members.
The second stage of the analysis involves estimating impacts for the predicted compliance subgroups and transforming these results into estimates of impacts on true compliance subgroup members.
Making this connection between predicted subgroup impacts and true subgroup impacts requires one of two strong assumptions:

\begin{assumption}[Constant individual level treatment effect]\label{assump:aspes_strong}
The impact of nonparticipation is the same for all nonparticipants. Also, the impact of participation is the same for all participants.
\end{assumption}

\begin{assumption}[Constant average treatment effect]\label{assump:aspes_weak} 
The mean impact of actual subgroup membership (for the participants group and for the nonparticipants group) is uncorrelated with the likelihood of predicted subgroup membership (and uncorrelated with the characteristics used to predict it).
\end{assumption}

Under Assumption~\ref{assump:aspes_weak}, if program participation is predicted based a handful of covariates, such as social economic status, gender, marital status, education level, and hours worked per week, then none of these covariates (or a weighted combination of these covariates) can be correlated with the mean impact of participating in the program.


\citet{Peck2003} claims that one can create a linear combination of multiple covariates to serve as a proxy variable that gives Assumptions~\ref{assump:aspes_strong} and~\ref{assump:aspes_weak}.
We believe that picking a single covariate, such as site, to act as such a proxy instrument (and then possibly adjusting for additional auxiliary covariates) is a more realistic way to obtain the strong identification conditions above.


\section{Simulation with correctly specified model}\label{sec:simple_sim}
To disentangle the effects of measurement error from model misspecification, we conducted a simulation study where the linear regression model is correctly specified.
We considered two active treatment cases under this `simple' simulation: one case where there is no compliance-impact confounder, and a second case where there is a compliance-impact confounder that affects $\ITT_{lc|k}$ and $\ITT_{hc|k}$ with different strengths.
For each case, we applied the same estimation methods as described in the main text (naive OLS, bootstrap OLS, no-measurement-error OLS, with or without confounder adjustment) to estimate the population average treatment impacts for Low-Quality Compliers, $\ITT_{lc}$, and for High-Quality Compliers, $\ITT_{hc}$.
Simulation code is available upon request.

\subsection{Relationships between confounder, relative proportion of High-Quality Compliers, and site-specific principal stratum impacts} \label{sec:simplesim_relationships}

\begin{figure}[H]
 \centering
  \includegraphics[scale=0.5]{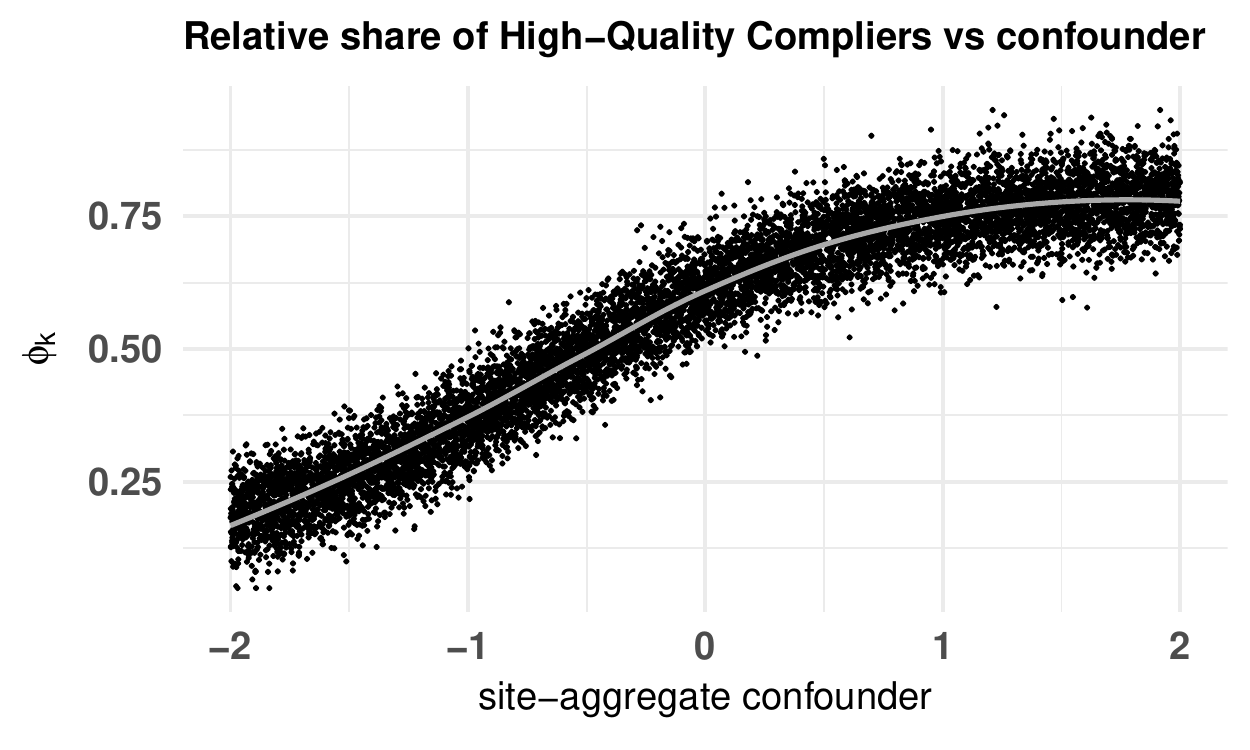}
 \caption{{\bf $\phi_k$ vs confounder} for 10,000 sites generated from `simple' dgp. There is a non-linear relationship between the confounder and the relative proportion of High-Quality Compliers in a site.}
 \label{fig:phi_vs_X_supp}
\end{figure}

\begin{figure}[H]
 \centering
  \includegraphics[scale=0.5]{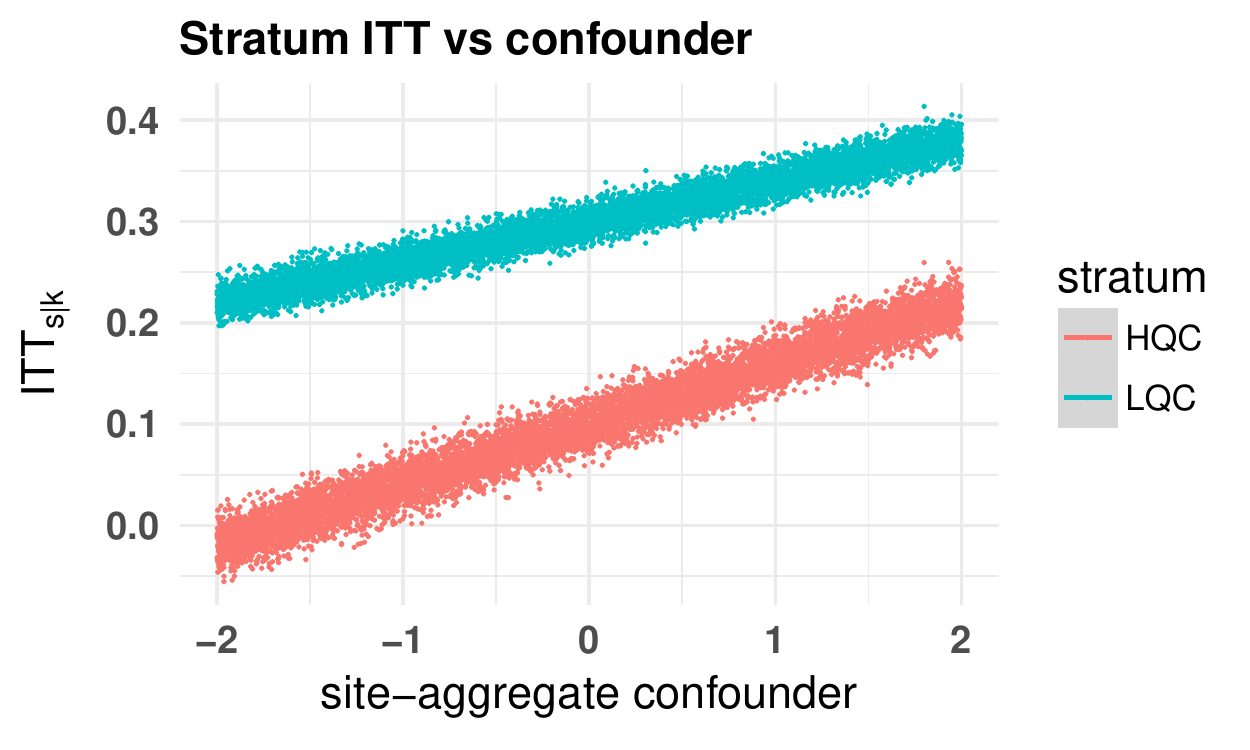}
 \caption{{\bf Stratum ITT vs confounder} for 10,000 sites generated from `simple' dgp. $\ITT_{lc|k}$ and $\ITT_{hc|k}$ are linear in the confounder $X_k$, with different slopes.}
 \label{fig:PCEs_vs_X_supp}
\end{figure}

We consider a uniformly distributed site-level covariate, $X_k$, that affects the proportion of Compliers in each site who are High-Quality, and separately, also affects the site-specific average treatment impacts for High-Quality and Low-Quality Compliers.
The site-specific treatment impacts for Low- and High-Quality Compliers are linear in $X_k$ (see Figure~\ref{fig:PCEs_vs_X_supp}).
The relationship between $\phi_k$ and $X_k$ is non-linear (see Figure~\ref{fig:phi_vs_X_supp}); however, when we correctly include $X_k \times \phi_k$ and $X_k \times (1-\phi_k)$ interaction terms in the OLS regression adjustment, we can unbiasedly recover the population average principal causal effects.

\subsection{Active treatment effects with no compliance-impact confounder}
\begin{figure}[H]
\begin{subfigure}{.6\textwidth}
  \centering
  \includegraphics[width=1\linewidth]{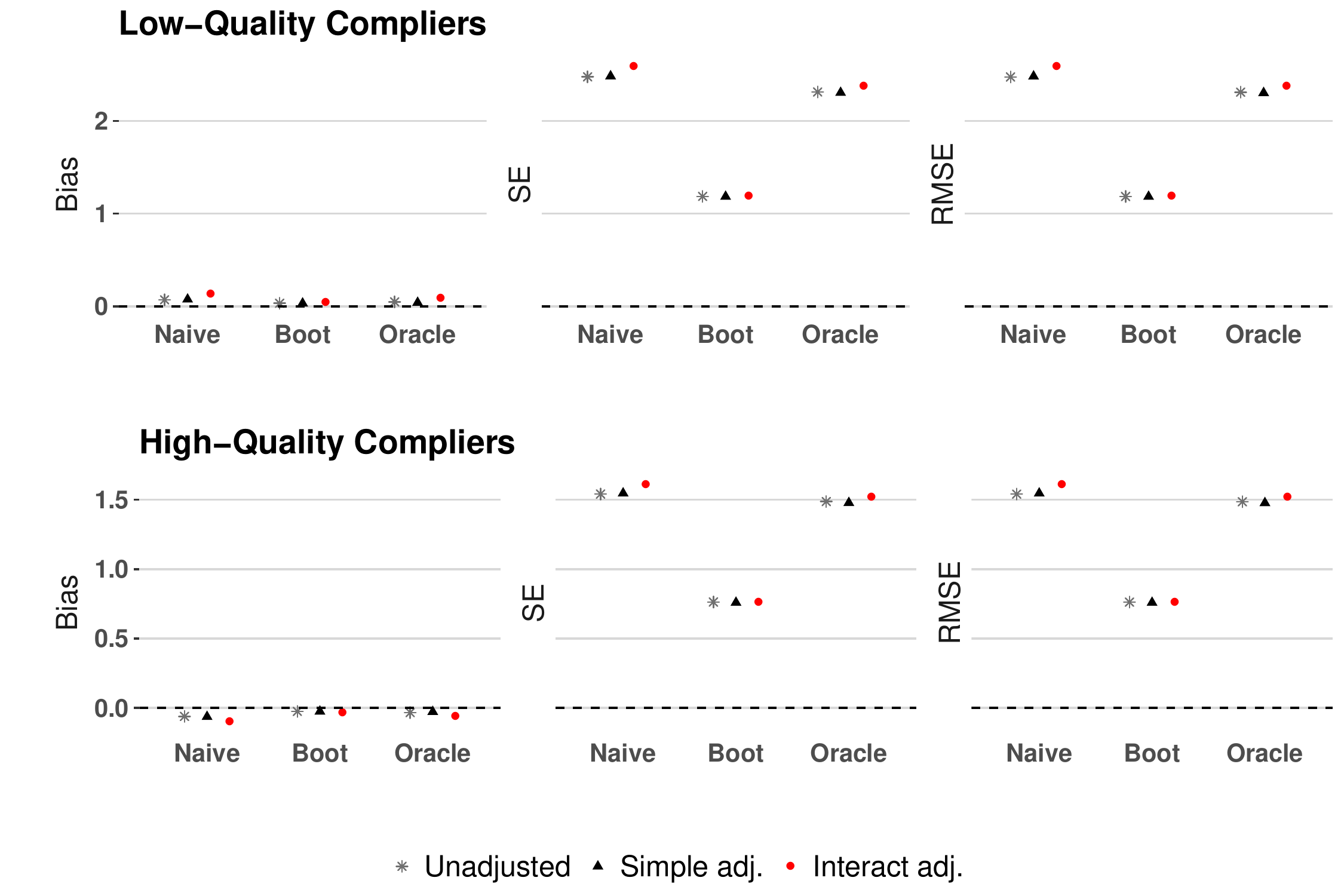}
  \caption{ Average bias, SE, and RMSE of each estimator. 
  }
  \label{fig:simplesim_noX_threestats}
\end{subfigure}\hspace{0.01\textwidth}
\begin{subfigure}{.37\textwidth}
  \centering
  \includegraphics[width=1\linewidth]{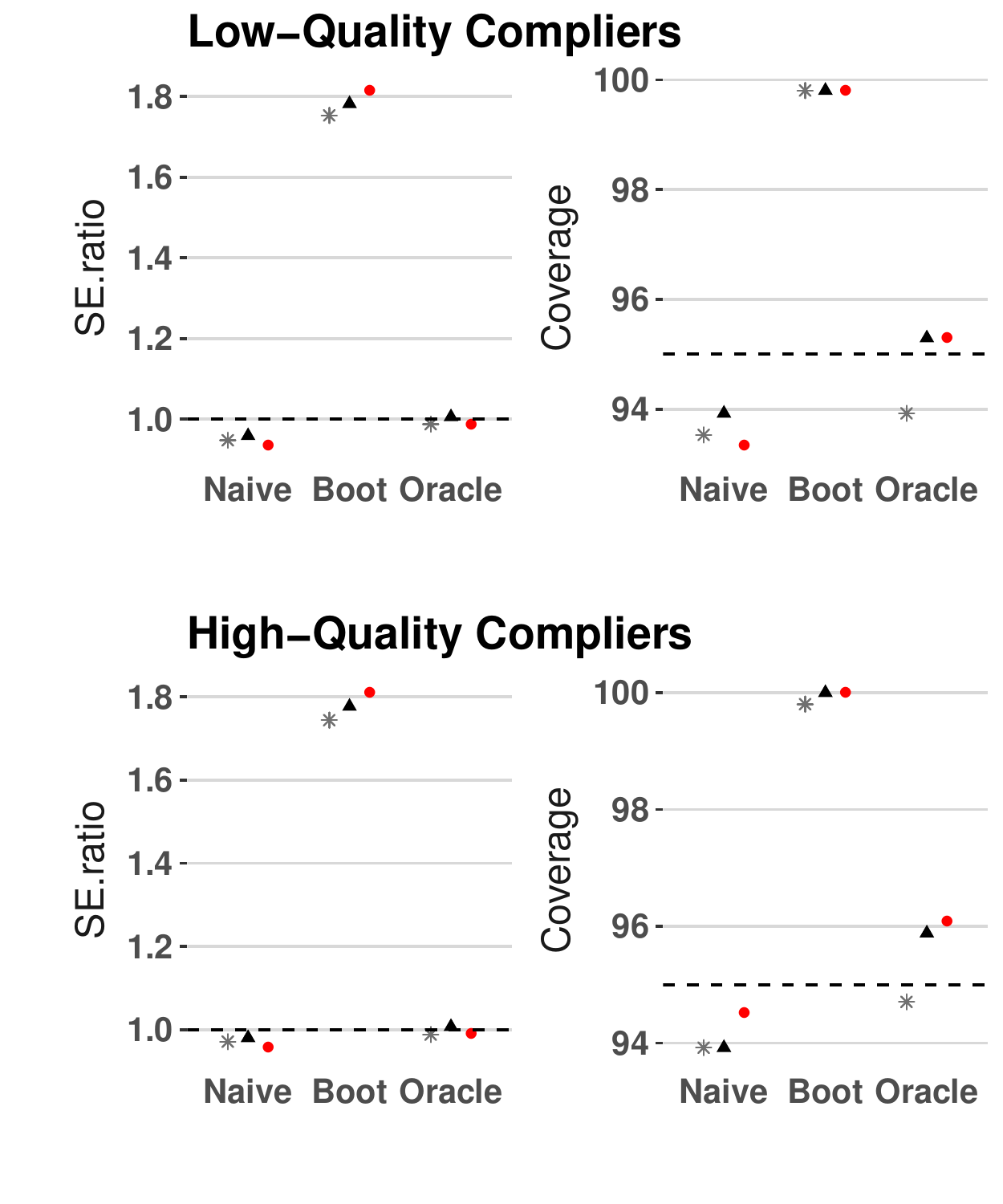}
  \caption{ Ratio of estimated to true SE of each estimator, and coverage of nominal 95\% CIs. 
  }
  \label{fig:simplesim_noX_cover}
\end{subfigure}
\caption{
          {\bf Simulation results under active treatment effects with no confounder.} 
          {\em Naive} ({\em Unadjusted}) denotes the the OLS regression coefficient estimators of Eq.~\eqref{eq:ols-unadjusted_supp}. {\em Boot} is the same OLS model fitted using case-resampling of students within sites to account for variation in site-level moment estimates. {\em Oracle} is the OLS model fitted to the true $\LATE_k$ and $\phi_k$, rather than estimated site-level quantities. Each estimation method is implemented without covariate adjustment, with a simple linear adjustment, and with an interaction term adjustment for the auxiliary covariate, $X$.
          }
\label{fig:simplesim_noX}
\end{figure}

When there are active stratum impacts but no compliance-impact confounder, the naive OLS model, bootstrap estimation method, and oracle model are all unbiased for $\ITT_{lc}$ and $\ITT_{hc}$ (see Figure~\ref{fig:simplesim_noX_threestats}).
The bootstrap method is the most stable method estimation method, but it overestimates the true SE, which leads to overly wide confidence intervals (see Figure~\ref{fig:simplesim_noX_cover}).

\subsection{Active treatment effects with one compliance-impact confounder}
\begin{figure}[H]
\begin{subfigure}{.6\textwidth}
  \centering
  \includegraphics[width=1\linewidth]{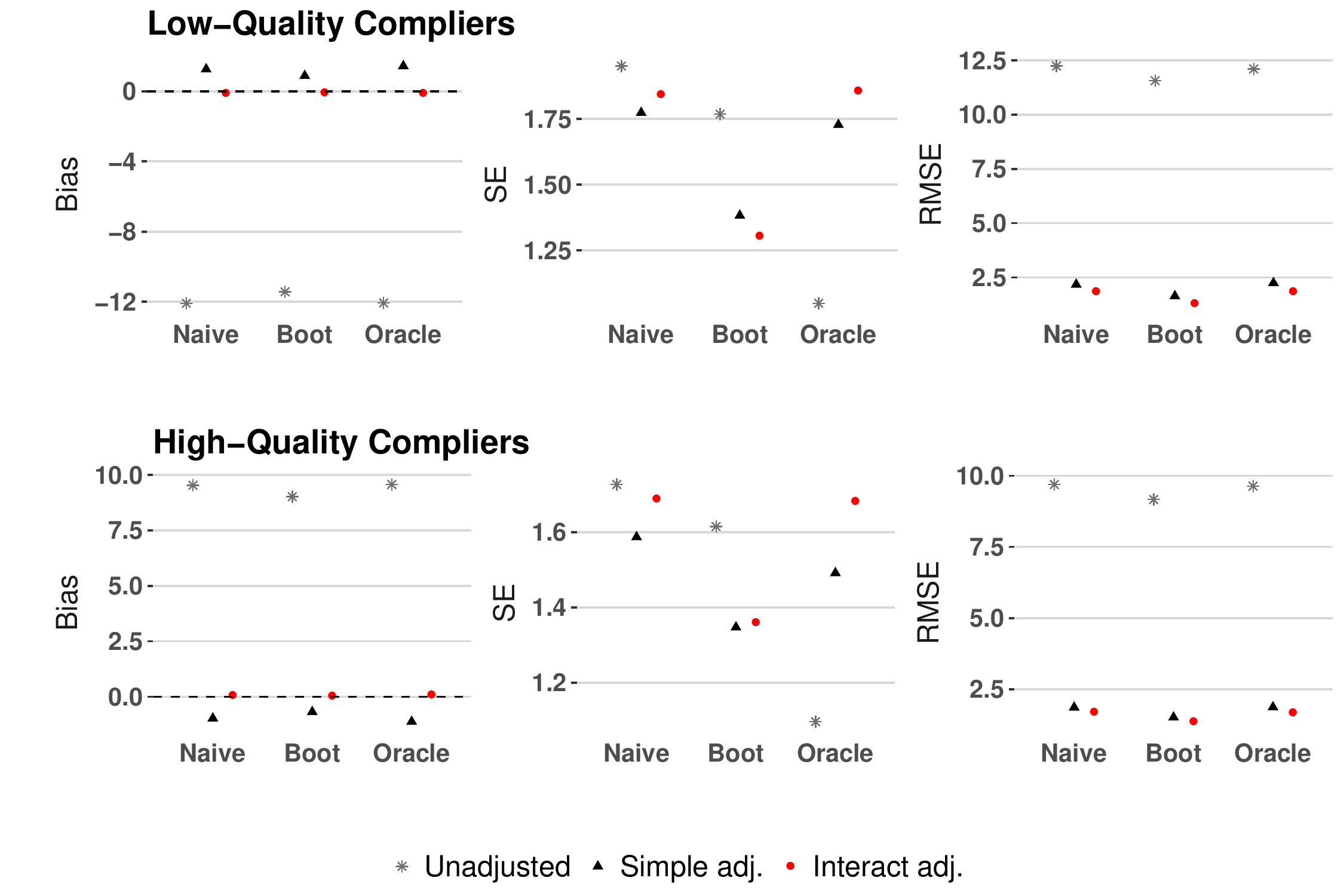}
  \caption{ Average bias, SE, and RMSE of each estimator. 
  }
  \label{fig:simplesim_withXv2_threestats}
\end{subfigure}\hspace{0.01\textwidth}
\begin{subfigure}{.37\textwidth}
  \centering
  \includegraphics[width=1\linewidth]{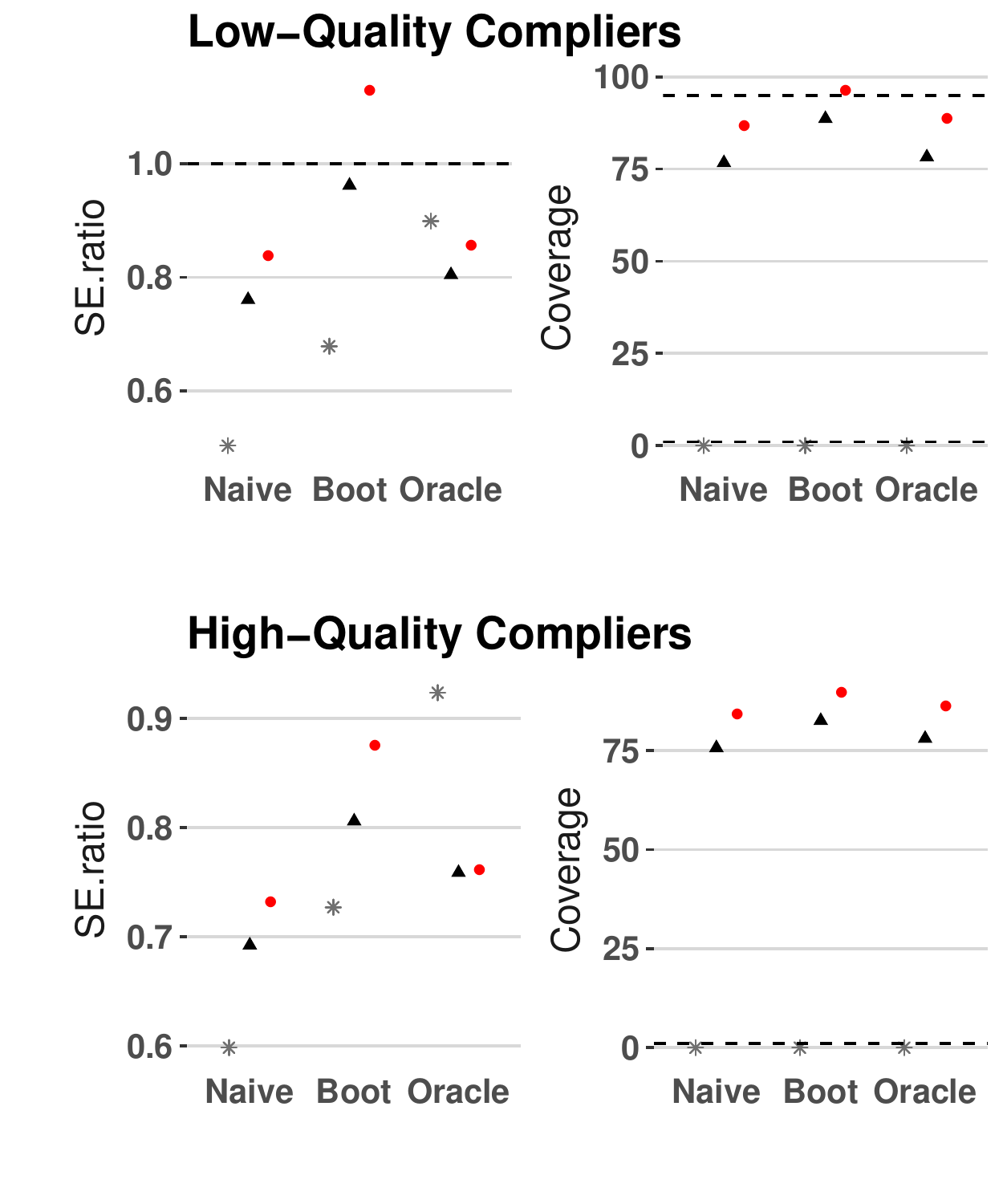}
  \caption{ Ratio of estimated to true SE of each estimator, and coverage of nominal 95\% CIs. 
  }
  \label{fig:simplesim_withXv2_cover}
\end{subfigure}
\caption{
          {\bf Simulation results under active treatment effects with one compliance-outcome confounder.} 
          Monte Carlo standard errors for Bias and RMSE are less than 0.1 percentage point; Monte Carlo standard errors for Coverage is less than one percentage point. {\em Naive} ({\em Unadjusted}) denotes the the OLS regression coefficient estimators of Eq.~\eqref{eq:ols-unadjusted_supp}. {\em Boot} is the same OLS model fitted using case-resampling of students within sites to account for variation in site-level moment estimates. {\em Oracle} is the OLS model fitted to the true $\LATE_k$ and $\phi_k$, rather than estimated site-level quantities. Each estimation method is implemented without covariate adjustment, with a simple linear adjustment, and with an interaction term adjustment for the auxiliary covariate, $X$.
          }
\label{fig:simplesim_withXv2}
\end{figure}

When there is a compliance-impact confounder that affects High-Quality and Low-Quality Compliers with different strengths (as described in Section~\ref{sec:simplesim_relationships}), a correctly specified linear model with interaction adjustment for the confounding variable (see Eq.~\ref{eq:ols-int-adjusted_supp})
leads to unbiased estimates of $\ITT_{lc}$ and $\ITT_{hc}$ under each estimation method, with the bootstrap estimator giving the most stable estimates (see Figure~\ref{fig:simplesim_withXv2_threestats}).
The bootstrap estimator with interaction adjustment generally has highest coverage, but depending on the simulation parameters specified (e.g., the strength of the confounder), none of the estimation methods are guaranteed to give confidence intervals with valid coverage (see Figure~\ref{fig:simplesim_withXv2_cover}).

\section{Calibrated simulation study} \label{sec:simulation}

We ran a simulation study to compare the estimation performance of the LATE versus ITT version of the OLS model (models~\ref{eq:ols-unadjusted_supp} and~\ref{eq:ITT_model_supp}, respectively), with and without auxiliary covariate adjustment (models~\ref{eq:ols-adjusted_supp} and~\ref{eq:ols-int-adjusted_supp}), and with and without bootstrap case-resampling to capture measurement error in the site-level quantities. 
We also included estimators where the site-level OLS model is fitted to true site-level statistics rather than estimated moments, to separate out issues of measurement error from the overall approach of conditioning on covariates to achieve zero site-level correlation.
The estimands of interest are the population average principal causal effects for Low- and High-Quality Compliers.

We calibrated our simulation design to remain close to the characteristics of the actual ECHS data.
In particular, we created student-level models for generating individuals of the different principal strata types, and then aggregated them to the site level as one would do in practice. 
This means that when we have a compliance-impact confounder, a site-level regression with a linear specification for the confounder is not necessarily correctly specified. 
To further validate our overall methods and investigate the impact of measurement error when the linear model specification is correct, we conducted an additional simulation, which we describe in Section~\ref{sec:simple_sim}.
There, we see that we are in fact able to remove all bias with correct model specification, even when the site-level statistics are estimated with noise.
We also show that bootstrap methods generate wider confidence intervals than the other estimation methods.

\subsection{Data generating process} \label{sec:sim_setup}

Our simulation takes on a modular form, roughly divided into four steps. 
We outline the data generating process (dgp) here.
Complete R scripts to generate data and estimate impacts are available upon request.


\begin{enumerate}
	\item \textbf{Sample sites.} 
	We randomly sample entire lottery sites with replacement from the ECHS student-level data, fixing the total number of sites at the originally observed $38$.
	We retain the site-aggregate covariate `eighth grade standardized reading score' to use for auxiliary covariate adjustment.
	Drawing from the empirical data allows us to generate hypothetical samples whose structure mimics that of the real experiment.

	\item \textbf{Generate potential outcome schedule.} 
	We consider two scenarios: with and without a site-level confounding variable. 
	For each sampled site, we randomly sample students' principal stratum membership based on the empirical distribution observed in the ECHS data, and, when applicable, the site-level confounder.
	We then generate potential outcomes under both treatment and control via a logistic regression, setting the probability that a student is on-track under control to 0.5. 
	The probability of a student being on-track under treatment is, on the logit scale, a linear function of her principal stratum membership and, when applicable, the site-level confounder.
	Importantly, since the dgp includes a logistic regression, a site-level regression with a linear specification for the confounder will be misspecified. 

	\item \textbf{Randomize students.} 
	Given the full potential outcomes schedule, we then randomize students to treatment or control, fixing the total number of treated and control students in each site to those originally observed in ECHS. This yields a hypothetical observed data set.

	\item \textbf{Estimate impacts.} 
	We estimate $\ITT_{hc}$ and $\ITT_{lc}$ by fitting different versions of the OLS models to site-aggregate quantities calculated from the hypothetical observed data.
\end{enumerate}

We repeat steps one through four many times\footnote{Each dgp was replicated 500 times.} to calculate the mean bias, standard error (SE), and root mean square error (RMSE) of each estimator, as well as the coverage of nominal 95\% confidence intervals (CIs).

\subsection{Results} \label{sec:sim_results}

We summarize findings from three dgp scenarios where we implement the various estimation methods described above.
We consider two forms of linear adjustment for average reading score in site: simple linear adjustment (Eq.~\ref{eq:ols-adjusted_supp}), and interaction adjustment (Eq.~\ref{eq:ols-int-adjusted_supp}).\footnote{All our regression adjustments are conducted using grand mean centered site-average reading score, wherein reading score is grand mean centered at the student level and then averaged by site.}
Throughout, we use the sample size weighted estimators of Eqs.~\eqref{eq:ITTlc_estimator_supp} and~\eqref{eq:ITThc_estimator_supp} 
to accommodate different site sizes.
We are targeting average effects across individuals, not across sites.
Because we found no significant difference in estimation performance of the LATE versus ITT version of the OLS models, 
we present here only results for the LATE estimation methods.

For each dgp, we describe the average bias, SE, and RMSE of the various estimators for $\ITT_{lc}$ and $\ITT_{hc}$.
We also investigate the conservativeness of each estimation method's SE estimator by looking at the ratio of each average SE estimate to its corresponding Monte Carlo (estimated) true SE.
This SE ratio helps explain the observed coverage of nominal 95\% confidence intervals.
All these statistics are expressed in percentage points. 
Monte Carlo standard errors are all less than one percentage point. 

\paragraph{Scenario 1: Null treatment effects with no compliance-impact confounder}
We evaluated a dgp where $\ITT_{lc} = 0$, $\ITT_{hc} = 0$, and there are no confounding variables.
In this case, there is no significant difference in mean bias nor variance between the unadjusted and reading score-adjusted versions of each estimation method.
All the estimation methods are unbiased for both principal causal effects.
The bootstrap estimators give valid confidence intervals and have smaller RMSE than the Naive OLS estimators that do not use bootstrap resampling to account for measurement error.

\begin{figure}[h!]
\begin{subfigure}{.6\textwidth}
  \centering
  \includegraphics[width=1\linewidth]{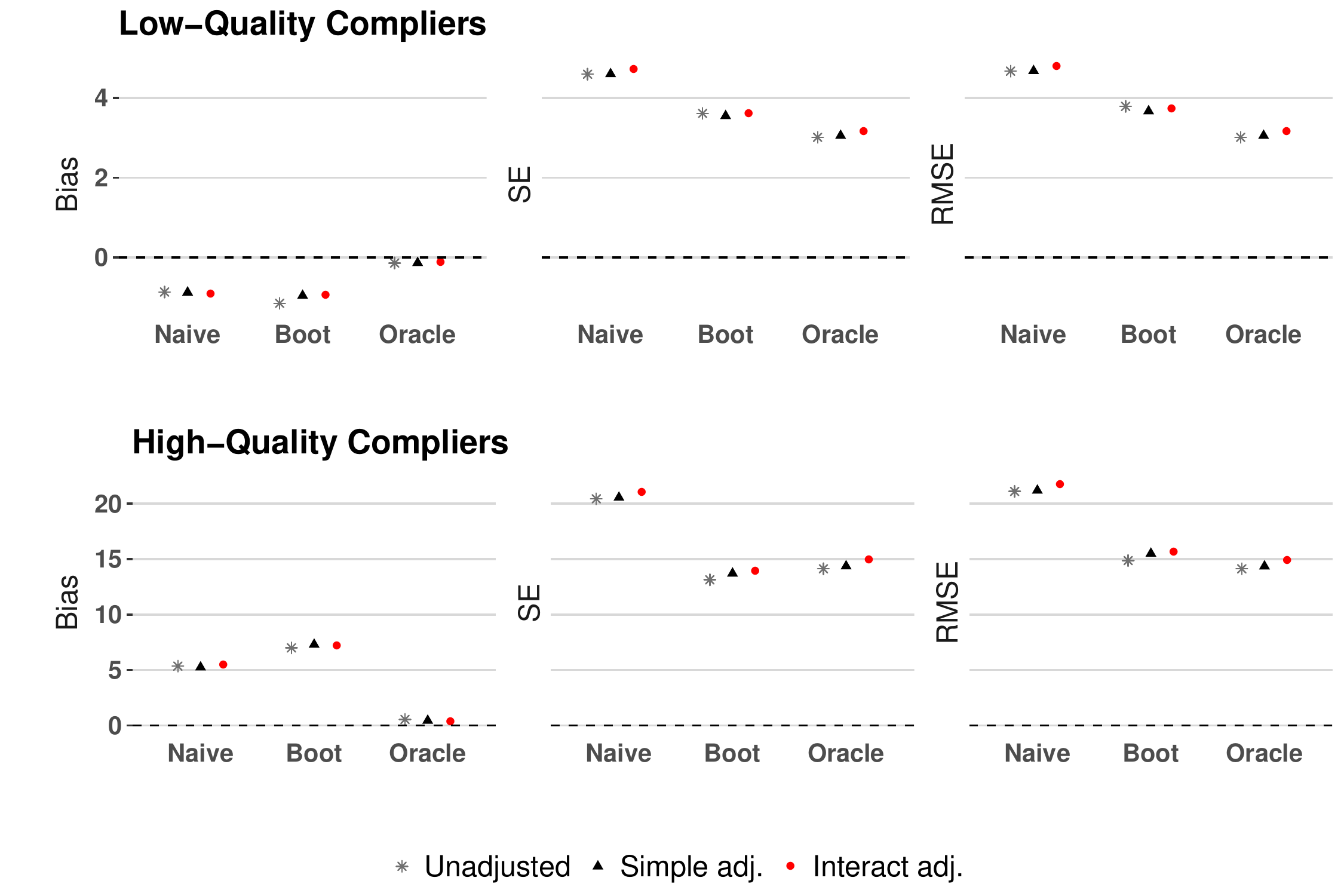}
 \caption{Average bias, SE, and RMSE of each estimator. 
 }
  \label{fig:activePCE_noconfounder_threestats}
\end{subfigure} \hspace{0.01\textwidth}
\begin{subfigure}{.37\textwidth}
  \centering
  \includegraphics[width=1\linewidth]{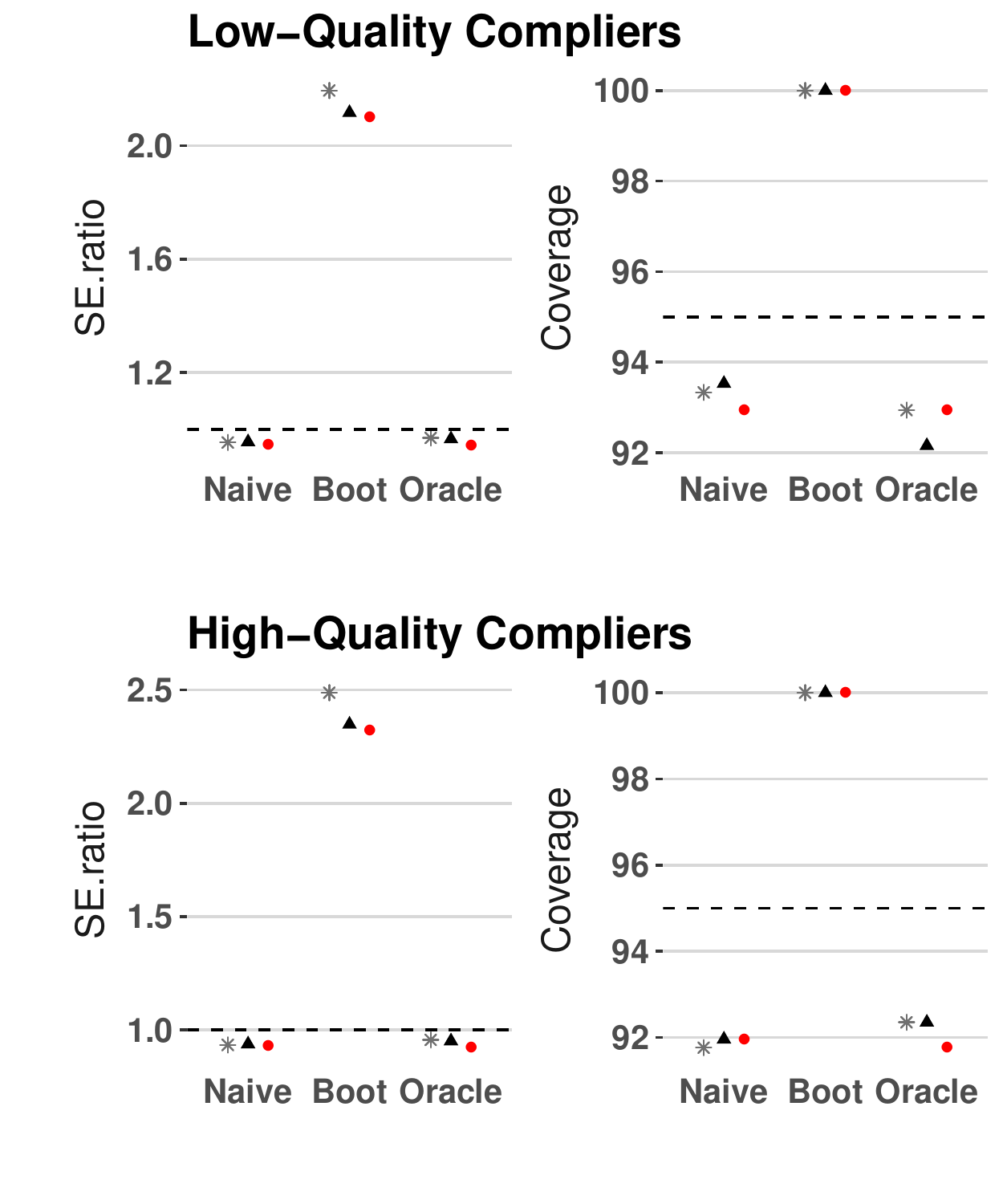}
  \caption{Ratio of estimated to true SE of each estimator, and coverage of nominal 95\% CIs.
  }
  \label{fig:activePCE_noconfounder_cover}
\end{subfigure}
\caption{{\bf Simulation results under active treatment effects with no compliance-impact confounder.} 
{\em Naive} ({\em Unadjusted}) denotes the the OLS regression coefficient estimators of Eq.~\eqref{eq:ols-unadjusted_supp}. {\em Boot} is the same OLS model fitted using case-resampling of students within sites to account for variation in site-level moment estimates. {\em Oracle} is the OLS model fitted to the true $\LATE_k$ and $\phi_k$, rather than estimated site-level quantities. Each estimation method is implemented without covariate adjustment, with a simple linear adjustment, and with an interaction term adjustment for site-average reading score.}
\label{fig:activePCE_noconfounder}
\end{figure}

\paragraph{Scenario 2: Active treatment effects with no compliance-impact confounder}
Here our dgp has $\ITT_{lc} = 12.32$ percentage points, $\ITT_{hc}= 0.08$ percentage points, and there are no confounding factors.

In this scenario, there is no significant difference between the unadjusted, simple reading score adjusted, or interaction adjusted versions of each estimation method, as measured by average bias, SE, RMSE, and coverage (see Figure~\ref{fig:activePCE_noconfounder}).
Second, the no-measurement-error (Oracle) estimators are unbiased for $\ITT_{lc}$ and $\ITT_{hc}$; however, their corresponding CIs, calculated using heteroskedastically-consistent (HC1) standard errors, are too narrow and undercover.
Third, the estimation methods that have site-level measurement error (Naive and Boot) are biased.
Bias under Boot is comparable or slightly worse than bias under Naive, although the bootstrap method has more stable estimates, as indicated by Boot's smaller SE and RMSE (see Figure~\ref{fig:activePCE_noconfounder_threestats}).
Lastly, because Boot overestimates the true standard errors by more than double their value, the coverage of Boot CIs is above the nominal 95\%.
We speculate that replacing HC1 standard errors with regular standard errors when using the bootstrap estimation procedures might lead to narrower yet valid CIs.

\paragraph{Scenario 3: Active treatment effects with a compliance-impact confounder}
We also investigate the performance properties of our various estimators for a dgp where $\ITT_{lc} = 10.9$ percentage points, $\ITT_{hc}= 2.1$ percentage points, and one pretreatment covariate, average reading score in site, confounds compliance type and the two Complier principal causal effects.
We specify site-average reading score to be a moderately strong confounder that affects the on-track status of High-Quality Compliers and Low-Quality Compliers with different strengths.
In particular, this dgp gives site-level pairwise correlations of 
$\text{Corr}\left(\phi_k,~read_k\right) = 0.69$, 
$\text{Corr}\left(\ITT_{lc|k},~read_k\right) = 0.58$, and
$\text{Corr}\left(\ITT_{hc|k},~read_k\right) = 0.13$.
The site-level covariance between $\phi_k$ and $\ITT_{lc|k}$ and $\ITT_{hc|k}$ are, respectively, approx 0.009 and 0.004. 
Note that our main identification results assume that these two covariances equal zero.
(Section~\ref{sec:sim_relationships} contains more details of the relationships between the confounder, $\phi_k$, and the treatment impacts.)

\begin{figure}[h!]
\begin{subfigure}{.6\textwidth}
  \centering
  \includegraphics[width=1\linewidth]{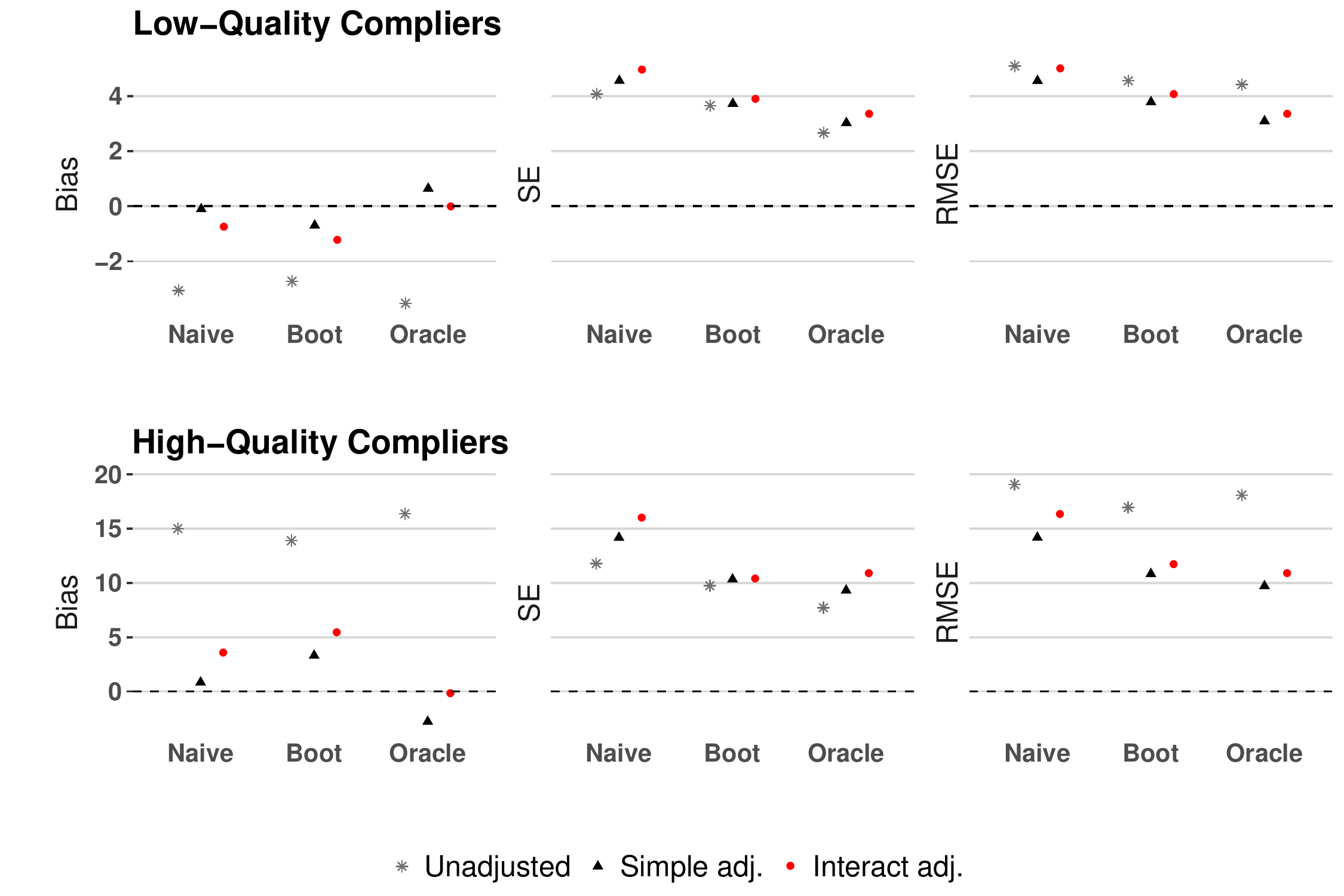}
  \caption{ Average bias, SE, and RMSE of each estimator. 
  }
  \label{fig:readconfounder_bX1half_threestats}
\end{subfigure}\hspace{0.01\textwidth}
\begin{subfigure}{.37\textwidth}
  \centering
  \includegraphics[width=1\linewidth]{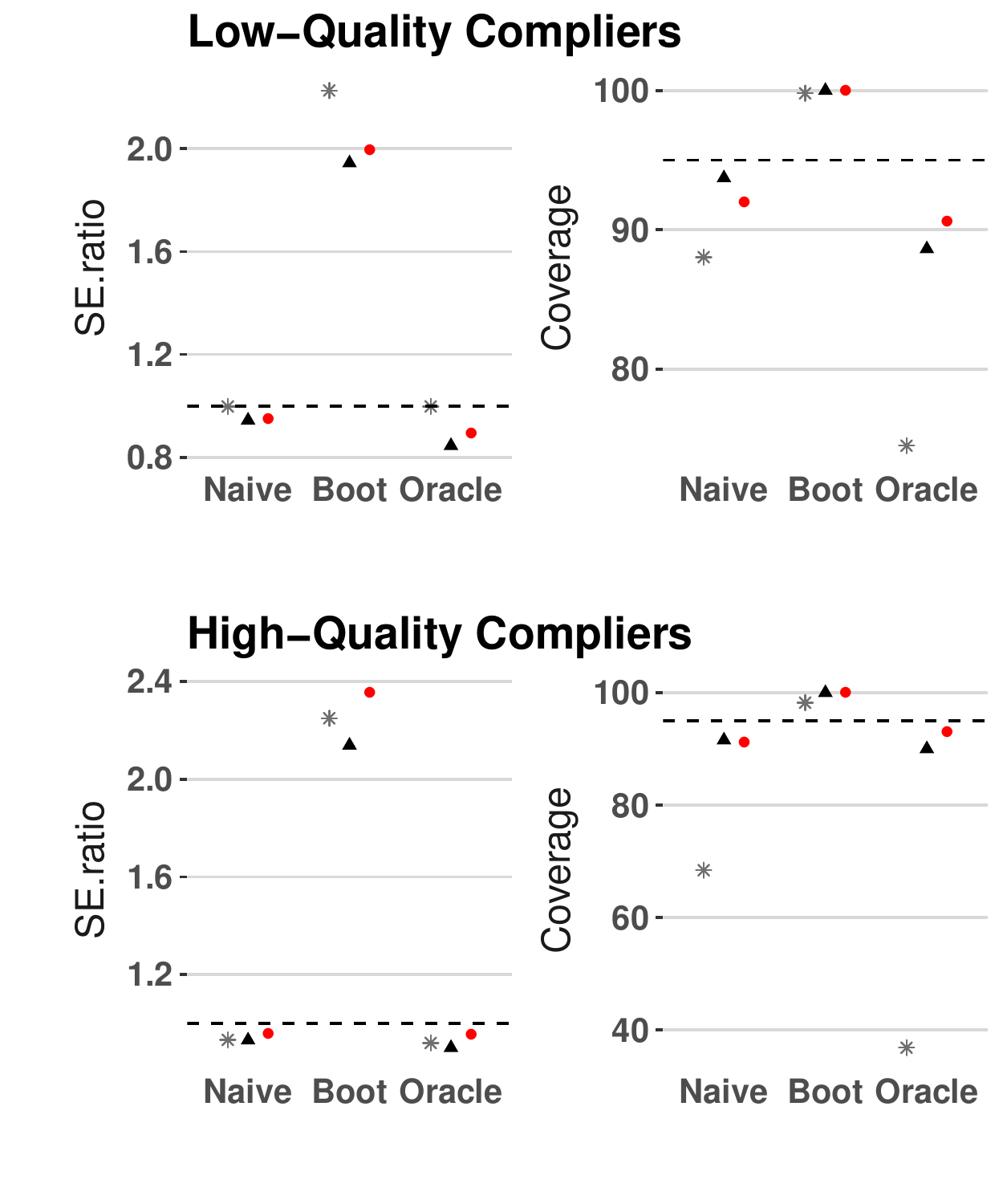}
  \caption{ Ratio of estimated to true SE of each estimator, and coverage of nominal 95\% CIs.
  }
  \label{fig:readconfounder_bX1half_cover}
\end{subfigure}
\caption{
          {\bf Simulation results under active treatment effects with one compliance-impact confounder.} 
          {\em Naive} ({\em Unadjusted}) denotes the the OLS regression coefficient estimators of Eq.~\eqref{eq:ols-unadjusted_supp}. {\em Boot} is the same OLS model fitted using case-resampling of students within sites to account for variation in site-level moment estimates. {\em Oracle} is the OLS model fitted to the true $\LATE_k$ and $\phi_k$, rather than estimated site-level quantities. Each estimation method is implemented without covariate adjustment, with a simple linear adjustment, and with an interaction term adjustment for site-average reading score.
          }
\label{fig:readconfounder_bX1half}
\end{figure}

When a compliance-impact confounder exists, not adjusting for the confounder can result in severely biased estimates of the principal causal effects (see {\em Unadjusted} estimates in Figure~\ref{fig:readconfounder_bX1half_threestats}).
The type of reading score adjustment that is sufficient to remove confounding bias depends on the correlation strengths between site-average reading score and the different Complier strata impacts, as well as the amount of nuisance measurement error in the site-level statistics.
Here, by setting reading score as a stronger confounder for Low-Quality Compliers than for High-Quality Compliers, interaction adjustment is needed to recover unbiased estimates of the principal causal effects when no measurement error exists; simple linear adjustment is insufficient (see Bias of Oracle model in Figure~\ref{fig:readconfounder_bX1half_threestats}).
Among the practical estimators with measurement error, the simple adjusted Naive estimator is unbiased while Boot is biased, but Boot has lower RMSE and higher coverage than Naive (Figure~\ref{fig:readconfounder_bX1half_cover}).
Furthermore, simple linear adjustment results in smaller average bias and higher coverage than interaction adjustment.
In this example, fitting a more complicated model with noisy data is not as good as fitting a simpler model with the same noisy data.

\paragraph{Takeaways from ECHS calibrated simulations}
Across various dgps, we find that the bootstrap estimator generally has comparable or slightly worse bias than the Naive estimator, but gives more stable estimates (smaller SE and RMSE) and valid, albeit often conservative, confidence intervals. 
When there is a compliance-impact confounder, adjusting for the confounder will provide less biased treatment effect estimates and better coverage.
Beyond that, which practical estimator and which form of covariate adjustment gives less biased estimates depends on the distribution of principal stratum membership across sites, the precision and accuracy with which we can estimate the site-level moments, and the strength of the confounder.
Confounder strength includes properties such as the magnitude of site-level pairwise correlations --- between the confounder and the relative share of High-Quality Compliers in each site, between the confounder and site-specific impacts for High-Quality Compliers, and between the confounder and site-specific impacts for Low-Quality Compliers --- as well as the functional relationship between the confounder and the principal causal effects.
For instance, if site-average reading score is correlated with compliance status, and if for any reading score, the treatment impact for Low-Quality Compliers is a constant number of percentage points higher than that for High-Quality Compliers, then a simple first-order linear adjustment of reading score will provide higher power for detecting a principal causal effect than further adjusting for a reading-score-by-compliance-status interaction.

\paragraph{Estimator performance under correctly specified model}
Section~\ref{sec:simple_sim} provides examples of simpler dgps where the linear model is correctly specified. 
There, we see that when there are active treatment effects for both Complier types but there is no compliance-impact confounder, all three estimation methods (Naive, Boot, and Oracle) are unbiased, while the bootstrap method is the most stable and has the smallest RMSE.
However, the bootstrap SE estimator is too conservative and gives nominal 95\% confidence intervals with 100\% coverage.
The Naive and Oracle CIs, on the other hand, have the correct coverage.
When there is a compliance-impact confounder that acts with different strengths on Low- and High-Quality compliers, the interaction adjusted model for each estimation method is unbiased while the unadjusted estimates are severely biased.
The interaction adjusted Boot estimator has a valid CI while the correctly specified Naive and Oracle estimators have confidence intervals that under-cover.

Again, as in the ECHS calibrated simulations, we find that the bootstrap estimation method is imperfect, but it generally performs as well as Naive in terms of bias, and better than Naive in terms of RMSE and coverage.

Consolidating results from both the calibrated and simpler simulations, we conclude that measurement error plus a misspecified model can lead to biased impact estimates. Measurement error alone, or model misspecification alone, however, do not guarantee bias.

\subsection{Relationships between reading score, relative proportion of High-Quality Compliers, and site-specific principal stratum impacts} \label{sec:sim_relationships}

\begin{figure}[H]
 \centering
  \includegraphics[scale=0.5]{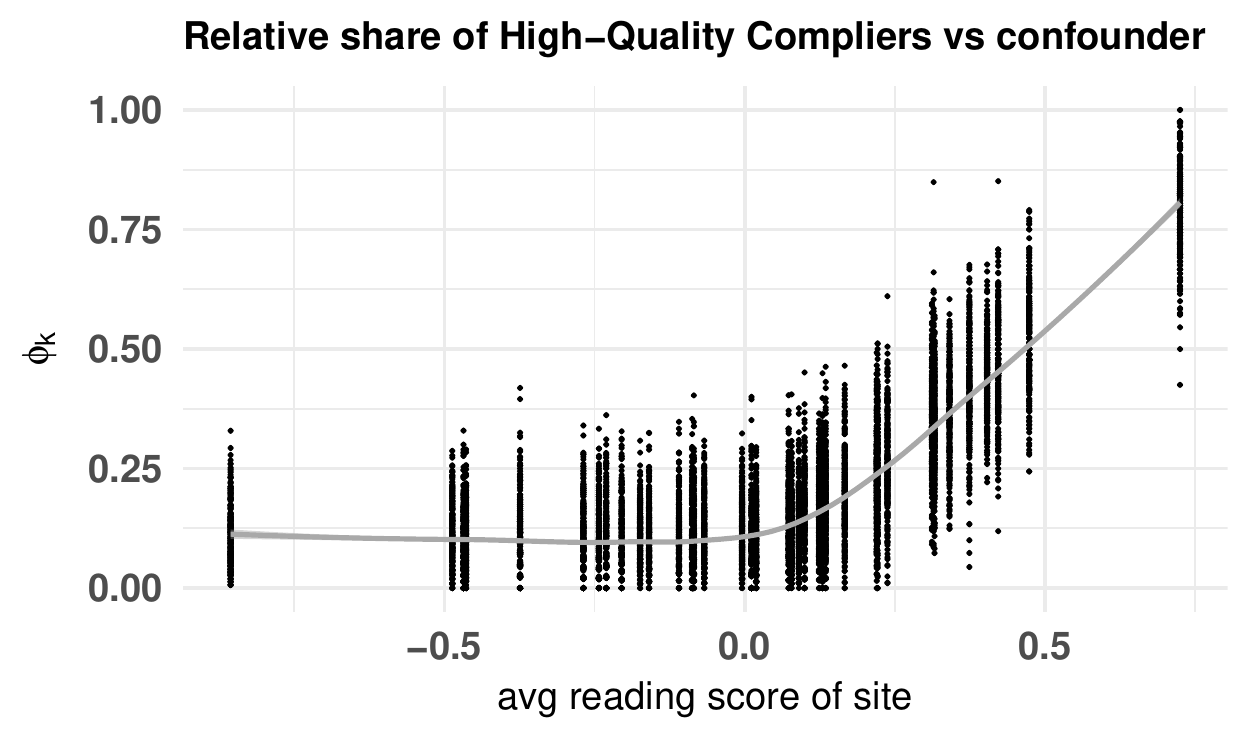}
 \caption{{\bf $\phi_k$ vs reading score} for 10,000 sites generated from calibrated dgp. There is a non-linear relationship between the confounder (site-average reading score) and the relative proportion of High-Quality Compliers in a site.}
 \label{fig:phi_vs_read_supp}
\end{figure}

\begin{figure}[H]
 \centering
  \includegraphics[scale=0.5]{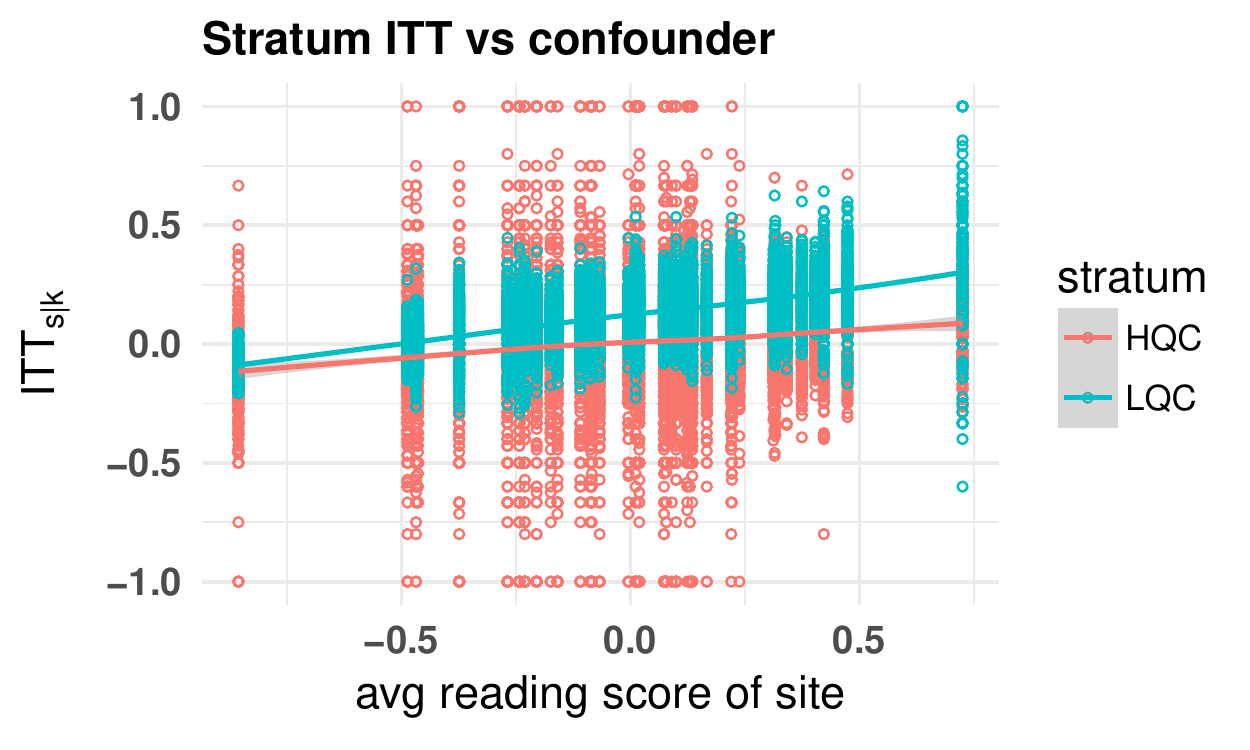}
 \caption{{\bf Stratum ITT vs reading score} for 10,000 sites generated from calibrated dgp. $\ITT_{lc|k}$ and $\ITT_{hc|k}$ vary differently as a function of reading score.}
 \label{fig:PCEs_vs_read_supp}
\end{figure}

With the calibrated simulation, we investigated a null treatment effects case and two active treatment impacts cases--one with a compliance-impact confounder and one without.
In the case with a compliance-impact confounder, site-average 8th grade reading score affects a student's principal stratum membership and, separately, her potential outcome under treatment.
At the site-level, this translates to the proportion of High-Quality Compliers in a site being correlated with the site-specific principal causal treatment effects.
In particular, there is a non-linear relationship between the eighth grade reading score and relative proportion of High-Quality Compliers (see Figure~\ref{fig:phi_vs_read_supp}), and $\ITT_{lc|k}$ and $\ITT_{hc|k}$ vary differently as a function of reading score (see Figure~\ref{fig:PCEs_vs_read_supp}).

\section{ECHS analysis using all 44 lotteries} \label{sec:echs_44lotteries}

\begin{figure}[h!]
 \centering
  \includegraphics[scale=0.5]{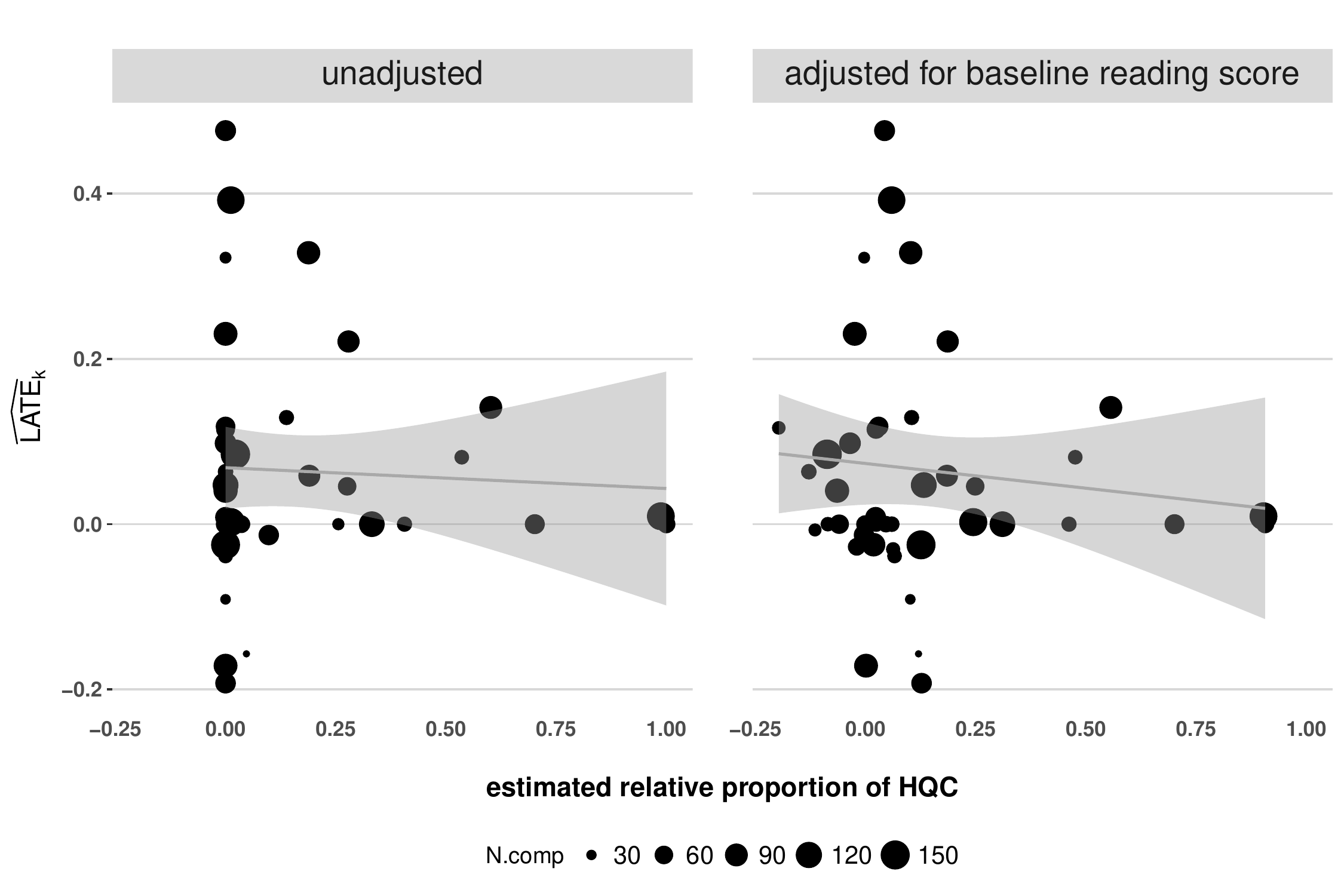}
 \caption{ {\bf ECHS site-level data for 44 lotteries.} Scatterplots of estimated site-specific Complier impacts (proportion on-track) versus (left panel) estimated relative proportion of High-Quality Compliers in a site, and (right panel) estimated \emph{residual} relative proportion of High-Quality Compliers in a site, after regressing $\hat{\phi}_k$ on eighth grade reading score. The size of the points indicate the number of Compliers in a site. The lines fit to the points correspond to linear regressions with a free intercept; the y-intercept for each line is an estimate for $\ITT_{lc}$, while the slope of each line is an estimate for the contrast $\ITT_{hc} - \ITT_{lc}$. The shaded grey regions are 95\% confidence intervals for the conditional mean outcome.} 
 \label{fig:LATE_vs_phihat_44lotteries}
\end{figure}

\begin{figure}[h!]
 \centering
  \includegraphics[scale=0.5]{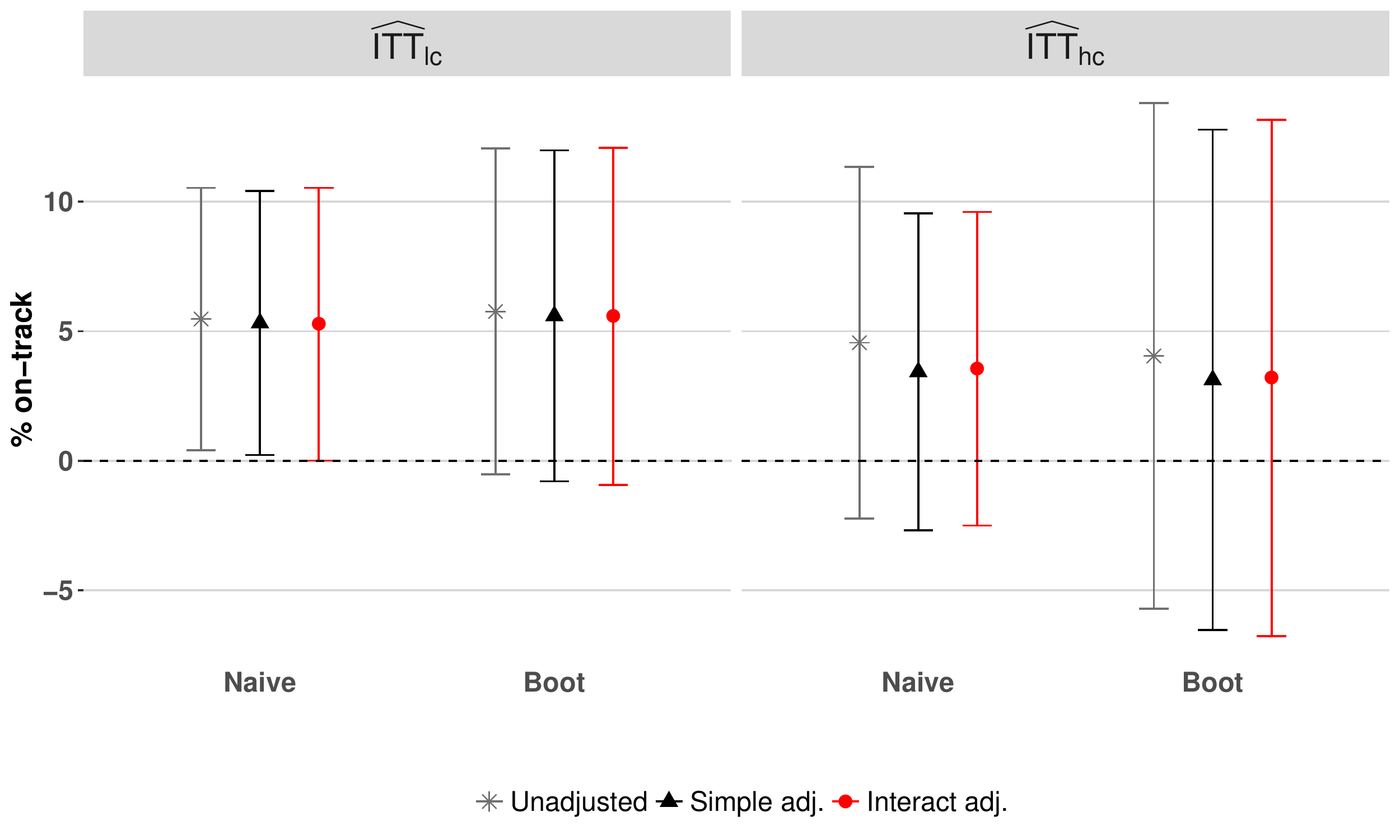}
 \caption{ {\bf Estimates of principal causal effects across 44 lotteries.} Point estimates and 95\% confidence intervals for Low- and High-Quality Complier principal causal effects are plotted for each estimation method.} 
 \label{fig:echs_estimates_LATEmod_44lotteries}
\end{figure}

\begin{figure}[h!]
 \centering
  \includegraphics[scale=0.5]{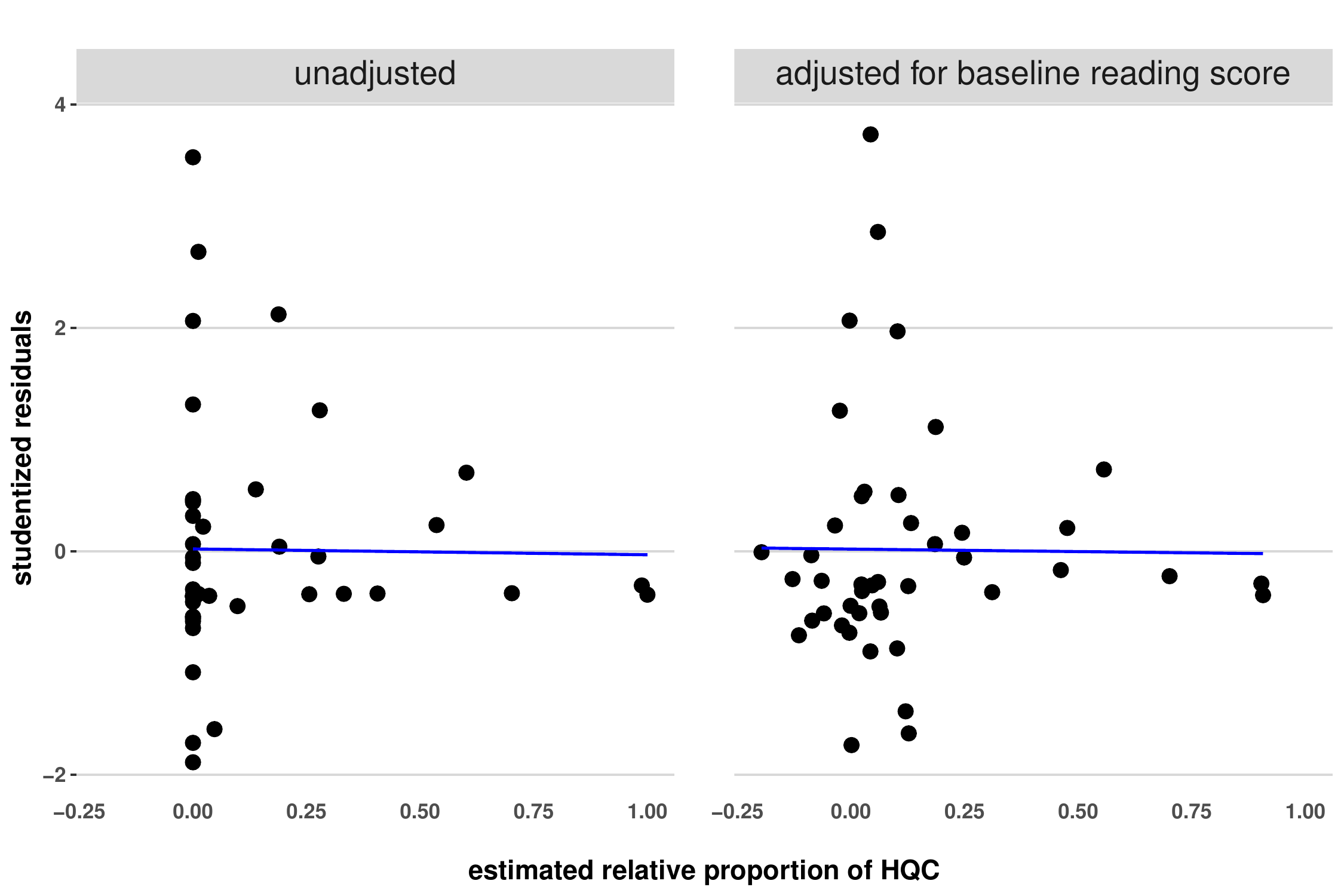}
 \caption{ {\bf Residual plots for 44 lotteries.} Studentized residuals versus estimated proportion of High-Quality Compliers for the Naive LATE model, where there is no baseline covariate adjustment (left panel) and where there is regression adjustment for eighth grade reading score (right panel). The blue lines are best-fit lines; one with a steep slope would indicate a violation of the (conditional) zero site-level correlation assumption needed to identify $\ITT_{lc}$ and $\ITT_{hc}$.}
 \label{fig:residuals_44lotteries}
\end{figure}

In the main text, we analyzed the impact of ECHS on the on-track status of students across 38 lotteries, excluding 6 lotteries in which all students were on-track at the end of their 9th grade year.
We performed the same analyses using all 44 lotteries (see data in Figure~\ref{fig:LATE_vs_phihat_44lotteries}) and found no difference in substantive results.
In particular, all point estimates for treatment impact are positive (Figure~\ref{fig:echs_estimates_LATEmod_44lotteries}), and we do not find meaningful differences in stratum impacts for High- vs Low-Quality Compliers.
However, partly because the Low-Quality Complier group is larger, we are more confident that the impact for this group is positive.
By contrast, the estimated impact for High-Quality Compliers is much noisier.
Finally, model checks do not indicate a violation of our zero correlation identifying assumptions (see Figure~\ref{fig:residuals_44lotteries}).

\end{document}